\documentclass[useAMS,usenatbib]{mn2e}
\usepackage[T1]{fontenc}
\usepackage{aecompl}
\usepackage{upgreek}

\usepackage[draft]{hyperref}

\usepackage{graphicx}
\usepackage{multirow}
\usepackage{amssymb}
\usepackage{hyperref}
\hypersetup{colorlinks=false,
		linkcolor=black,
		urlcolor=cyan}

\title[slow-evolving SLSNe I]{Complexity in the light curves and spectra of slow-evolving superluminous supernovae}
\author[C. Inserra et al.]{C. Inserra$^{1,2}$\thanks{E-mail: c.inserra@qub.ac.uk (CI)},
M. Nicholl$^{3}$,
T.-W. Chen$^{4}$, A. Jerkstrand$^{5}$, S. J. Smartt$^{1}$, T. Kr\"{u}hler$^{4}$,
\newauthor J. P. Anderson$^{6}$, C. Baltay$^{7}$, M. Della Valle$^{8}$,  M. Fraser$^{9}$, A. Gal-Yam$^{10}$, L. Galbany$^{11}$,
\newauthor E. Kankare$^{1}$, K. Maguire$^{1}$, D. Rabinowitz$^{7}$, K. Smith$^{1}$, S. Valenti$^{12}$ and D. R. Young$^{1}$
\\
$^{1}$Astrophysics Research Centre, School of Mathematics and Physics, Queen's University
  Belfast, Belfast BT7 1NN, UK\\
$^{2}$Department of Physics \& Astronomy, University of Southampton, Southampton, Hampshire, SO17 1BJ, UK\\
$^{3}$Harvard-Smithsonian Center for Astrophysics, 60 Garden Street, Cambridge, Massachusetts 02138, USA\\
$^{4}$Max-Planck-Institut f{\"u}r Extraterrestrische Physik, Giessenbachstra\ss e 1, 85748, Garching, Germany\\
$^{5}$Max Planck Institute for Astrophysics, Garching, Karl-Schwarzschild-Str. 1, Postfach 1317, D-85741 Garching, Germany\\
$^{6}$European Southern Observatory, Alonso de C\'ordova 3107, Vitacura, Casilla 190001, Santiago, Chile\\
$^{7}$Physics Department, Yale university, New Haven, CT 06511, USA\\
$^{8}$Capodimonte Observatory, Salita Moiariello 16, I-80131, Napoli, Italy\\
$^{9}$School of Physics, O'Brien Centre for Science North, University College Dublin, Belfield, Dublin 4, Ireland\\
$^{10}$Benoziyo Center for Astrophysics, Weizmann Institute of Science, 76100 Rehovot, Israel\\
$^{11}$Physics and Astronomy Department, University of Pittsburgh, Pittsburgh, PA 15260, USA\\
$^{12}$Dept. of Physics, University of California Davis, 1 Shields Ave, Davis, CA 95616, USA
}

\def\kms{km\,s$^{-1}$}
\def\Ha{H{$\alpha$}}
\def\Hb{H{$\beta$}}

\def\ni{$^{56}$Ni}
\def\co{$^{56}$Co}
\def\fe{$^{56}$Fe}

\def\M{M$_{\odot}$}

\def\an{LSQ14an}
\def\oiii{[O~{\sc iii}]}
\def\sii{[S~{\sc ii}]}
\def\siii{[S~{\sc iii}]}
\def\nii{[N~{\sc ii}]}
\def\neiii{[Ne~{\sc iii}]}
\def\ariii{[Ar~{\sc iii}]}
\def\nii{[N~{\sc ii}]}
\def\oii{[O~{\sc ii}]}
\def\oi{[O~{\sc i}]}

\begin{document}

\date{Received.....; accepted...........}

\pagerange{\pageref{firstpage}--\pageref{lastpage}} \pubyear{}

\maketitle

\label{firstpage}

\begin{abstract}
A small group of the newly discovered superluminous supernovae show broad and slowly evolving light curves. 
Here we present  extensive observational data for the slow-evolving superluminous supernova
LSQ14an, which brings this group of transients to four in total in the low redshift Universe (z$<$0.2; SN~2007bi, PTF12dam, SN~2015bn). 
We particularly focus on the optical and near-infrared evolution during the 
period from 50 days up to 400 days from peak, showing that  they are all fairly similar in their light curve and spectral evolution. 
LSQ14an 
shows broad, blue-shifted [O~{\sc iii}] $\lambda\lambda$4959, 5007 lines, as well as a blue-shifted [O~{\sc ii}] $\lambda\lambda$7320, 7330 and [Ca~{\sc ii}] $\lambda\lambda$7291, 7323. 
Furthermore, the sample of these four objects shows common features. 
Semi-forbidden and forbidden emission lines appear surprisingly early at 50--70 days and remain visible with almost no variation up to 400 days.  The spectra remain blue out to 400 days. There 
are small, but discernible light curve fluctuations in all of them. 
 The light curve of each shows a faster decline than \co\/ after 150 days and it further steepens after 300 days.  We also expand our analysis presenting X-ray limits for LSQ14an and SN~2015bn and discuss their diagnostic power. 
 These features are quite distinct from the 
faster evolving superluminous supernovae and are not easily explained in terms of only a variation in ejecta mass. 
While a central engine is still the most likely luminosity source, it appears that the ejecta structure is complex, with multiple emitting zones and {\it at least} some interaction between the 
expanding ejecta and surrounding material. 
 \end{abstract}


\section{Introduction}\label{sec:intro}
A new class of intrinsically bright supernovae, labelled superluminous supernovae \citep[SLSNe;][]{qu11,gy12}, was unveiled during last decade. They were originally identified as those supernovae (SNe) showing M$\sim-21$ mag \citep[e.g.][]{smi07,pa10,qu11,cho11,gy12,in13,how13} and divided in three different types \citep[see the early review of][]{gy12}. They now tend to be separated and classified based on their spectrophotometric behaviour rather than a simple magnitude threshold, 
 \citep[e.g.][]{pap15,in16b,pr16,lu16}. 
Those exhibiting hydrogen-free ejecta are grouped together as SLSNe~I, while those showing hydrogen as SLSNe~II. 

SLSNe~I, although intrinsically rare \citep{qu13,mc15,pr16} are the most commonly studied and usually found in dwarf, metal poor galaxies \citep[e.g.][]{lu14,le15,angus16,pe16,ch16}.  At about 30 days after peak, these SNe tend to show strong resemblances to normal or broad-lined type Ic SNe at peak luminosity \citep[e.g.][]{pa10,lm16} and thus are also usually referred to as SLSNe Ic \citep[e.g.][]{in13,ni13}. Their light curves show significant differences in width and temporal evolution
with the rise times being correlated with the declines rates \citep{ni15}. 
Whether or not there are two distinct groups, and two different physical mechanisms 
for the fast-evolving (e.g. SN~2010gx) and slow-evolving (e.g. SN~2007bi) 
explosions remains to be seen.

In addition, thanks to modern imaging surveys we have been able to differentiate between two groups of hydrogen rich SLSNe, which are  rarer by volume than SLSNe~I. The first is that of strongly interacting SNe such as SN~2006gy \citep{smi07,of07,ag09} of which the enormous luminosity is undeniably powered by the interaction of supernova ejecta with dense circumstellar (CSM) shells and hence usually referred to as SLSNe IIn to highlight the similarity with normal type IIn. Whereas, the second group is that of bright, broad lined H-rich SNe labelled SLSNe II, such as SN~2008es \citep{ge09,mi09}, which have shallow but persistent and broad Balmer lines \citep[see][for a review]{in16b}. They show spectroscopic evolution that is similar, in some ways to type II-L supernovae, but the evolution happens on a longer timescale. 

The mechanisms responsible to power the luminosities of SLSNe I \& II is still debated. However, the most favoured is that of a magnetar as a central engine, depositing additional energy into the supernova ejecta \citep{wo10,kb10,de12}. The popularity of this scenario is due to its ability to reproduce a variety of observables for all the types \citep[e.g.][]{in13,ni13,in16b}. However, alternative scenarios such as the accretion onto a central black hole \citep{dk13}, the interaction with a dense circumstellar medium \citep[e.g.][]{ci11} or between two massive shells \citep[pulsational pair instability,][]{woo07}, and a pair instability explosion \citep[e.g.][]{kb15} are still feasible alternatives.

Due to their intrinsic brightness, SLSNe~I are emerging as high redshift probes and might be a new class of distance indicators \citep{in14,pap15,sco16}. To use them with the new generation of telescopes such as the Large Synoptic Survey Telescope (LSST) or Euclid we need novel approaches  to better understand their progenitor scenario, reveal the mechanism powering the light curve and investigate on the nature of their differences. These new approaches include analyses of large dataset \citep[e.g.][]{ni15}, spectropolarimetric studies \citep{le15b,in16b,in16c} or devoted spectroscopic modelling of photospheric \citep{maz16} and nebular spectroscopy \citep{je16}.

The slowly evolving group of SLSNe I, of which the first to be discovered was SN~2007bi \citep{gy09,yo10}, is small 
in number but of particular interest because the light curves decline at a rate consistent with powering by 
$^{56}$Co during first 100-200 days after maximum. A  very large amount of explosively produced 
$^{56}$Ni is required to power the peak through $^{56}$Co decay (to produce stable $^{56}$Fe). With estimates 
in the region of 4-6M$_{\odot}$ this led  \citet{gy09} to propose a pair-instability explosion origin for SN~2007bi
(in combination with spectral analysis) from a very massive progenitor star above $M_{\rm ZAMS} \simeq$150M$_{\odot}$. 
Hence any new objects with similar light curve and spectra behaviour are of interest to study over long durations. 

Here we present an extensive dataset for  \an\/, a slow-evolving SLSNe~I discovered by the La Silla QUEST survey and 
studied by the Public ESO Spectroscopic Survey for Transient Objects \citep[PESSTO\footnote{www.pessto.org},][]{2015A&A...579A..40S}.   To draw a meaningful comparison, we  compare it with a sample of slow-evolving SLSNe~I
at low redshift (z$<$0.2).  There are three objects that share very similar characteristics that make up this comparison  -  SN~2007bi \citep{gy09,yo10}, PTF12dam \citep{ni13,ch15,vr16}, SN~2015bn \citep{ni16a,ni16b}. These objects all have 
 rest-frame spectra coverage up to at least 8000~\AA\/ and late time spectroscopy and photometry up to 400 days. 
We will also discuss the higher redshift, and perhaps more extreme PS1-14bj \citep{lu16}. These comparisons 
allow us to further investigate the explosion mechanism and  the progenitors  of these transients, as well as the role of  interaction with a circumstellar medium.

\section{LSQ14an Observations and data reduction}\label{sec:data}

\an\/ was discovered on {\sc ut} 2014 January 02.3 by the La Silla-Quest (LSQ) survey 
\citep{2013PASP..125..683B}. 
The LSQ camera used a broad composite filter covering the SDSS $g$ and $r$ bands,  from 4000~\AA\/ to 7000~\AA\/ and the pipeline initially reported a discovery 
magnitude of  $m\sim 19.1$ mag in this filter. The object was promptly classified within 24hrs by PESSTO as a SLSN roughly 50 days after peak brightness at $z\sim0.16$ with similarities to the slow-evolving PTF12dam and SN~2007bi \citep{atel}.  A more careful photometric analysis and calibration of the images from the QUEST camera to SDSS $r-$band, resulted in an initial magnitude of {\it r}=18.60$\pm$0.08 mag, as reported in Table~\ref{table:snm} (see Section \ref{sec:lc} for details of the photometry). 

\begin{figure}
\includegraphics[width=\columnwidth]{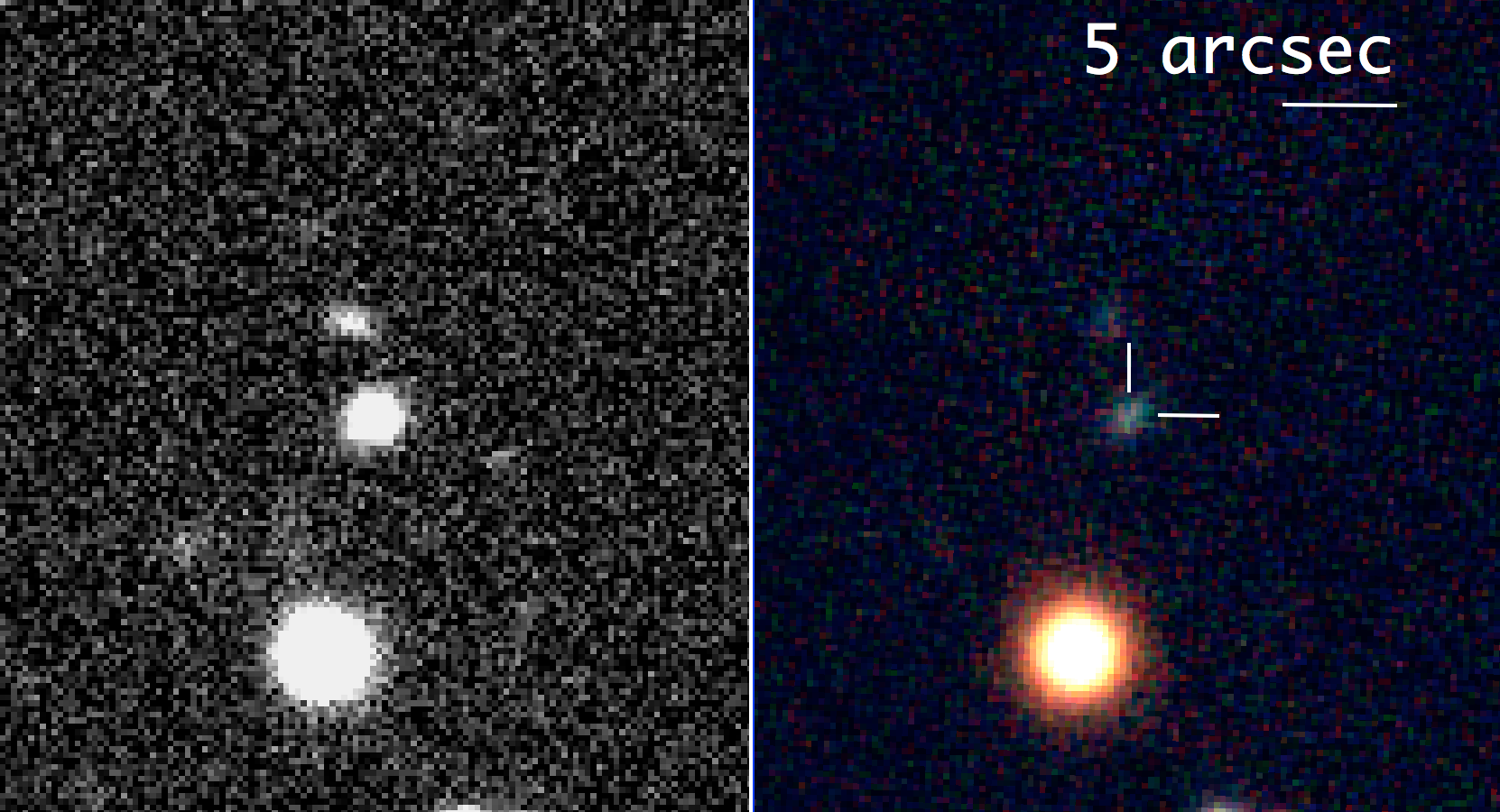}
\caption{The left panel shows the $r-$band EFOSC2 image of LSQ14an taken on 2014 Jan 31.
The right panel shows the colour composite of the host galaxy from the Pan-STARRS1 survey (constructed
from $g_{\rm P1}r_{\rm P1}z_{\rm P1}$ filters). The two images are astrometrically
aligned, with the centroid position of LSQ14an marked with the white cross
hairs on the right. This illustrates that the SN is coincident with the dwarf
galaxy host and that spectra of LSQ14an will be contaminated by the host galaxy
flux since it is unresolved. The image quality of the EFOSC2 image is 0.9\arcsec\/ FWHM and the PS1 $r_{\rm P1}$ image is 1.2\arcsec.
North is up and east to the left.}
\label{fig:host}
\end{figure}

Optical spectrophotometric follow-up was  obtained with the New Technology Telescope (NTT)+EFOSC2 as part of PESSTO and supplemented with 
other facilities. Photometry with EFOSC2 was obtained using {\it BVR+gri} filters and was augmented with observations from the Liverpool Telescope (LT)+IO:O in  $ gri$ filters.   
The EFOSC2 images were reduced (trimmed, bias subtracted and flat-fielded) using the PESSTO pipeline \citep{2015A&A...579A..40S}, while the LT images were reduced automatically by the LT instrument-specific pipeline. One epoch of 
near-infrared (NIR) imaging was obtained with the NTT and SOn oF Isaac (SOFI) instrument.
These data were also reduced using the PESSTO pipeline. To remove the 
contribution of the host galaxy to the SN photometry (see Section\,\ref{sec:host} for a data
on the host) we  applied a template subtraction technique \citep[with the {\sc hotpants}\footnote{http://www.astro.washington.edu/users/becker/hotpants.html} package based on the algorithm presented in][]{al00}. We used EFOSC2 $BVR$ images taken on 2015 April 17 
(MJD=57129, or an estimated +460d after peak), 
and the PS1 $gri$ images taken before 2013 June 24 (see Section~\ref{sec:host}) as the reference templates. On the other hand, we did not use any $JHK$ templates for our single epoch of observations since the contamination is on the order of a few per cent and has no significant effect on our analysis. Since instruments with different passbands were used for the follow-up of \an\/ we applied a passband correction \citep[P-correction,][]{in16b}, which is similar to the S-correction \citep{st02,pi04}, and allows  photometry to be standardised to a common system, which in our case is SDSS AB mags for {\it gri} and Bessell Vega mags for {\it BVR}. This procedure takes into account the filter transmission function and the quantum efficiency of the detector, as well as the intrinsic spectrum of \an, but does {\it not} include the reflectance or transmission of other optical components, as this is relatively flat across the optical range \citep[see][for further details]{in16b}.
Optical and NIR photometry are reported in Tables~\ref{table:snm}~\&~\ref{table:snmv} together with the coordinates and magnitudes of the sequence stars used (see Table~\ref{table:sref}).

Optical and near-infrared spectra were taken during the PESSTO follow-up campaign and 
were supplemented with two spectra from X-Shooter on the ESO Very Large Telescope (VLT). 
The log of spectral observations (eight separate epochs), with resolutions and wavelength ranges is given in Table\,\ref{table:sp}. Two further spectra were taken with the VLT at later epochs of (+365 and +410d) when the SN was in the nebular 
phase and are analysed in detail in \cite{je16}. The PESSTO spectra were reduced
in standard fashion as described in \cite{2015A&A...579A..40S}. 
The X-Shooter spectra were reduced in the same manner as described for the nebular 
spectra in \citet{je16}. We used the custom made pipeline described in \citet{2015A&A...581A.125K} which takes the ESO pipeline produced 2D spectral products 
\citep[][]{2010SPIE.7737E..28M} and uses  optimal extraction with a Moffat profile fit.  This pipeline produced flux calibrated spectra with rebinned dispersions of 0.4 \AA\,pix$^{-1}$ in the UVB+VIS arms and  0.6 \AA\,pix$^{-1}$ in the NIR arm. 
The PESSTO spectra taken up to April 2014 are available from the public data release
Spectroscopic Survey Data Release 2 (SSDR2) at ESO. Steps to retrieve the data 
from the ESO Science Archive Facility are available on the PESSTO website. The final 
EFOSC2 spectrum on 21 August 2014 will be released in the upcoming SSDR3. All spectra
will also be available on WISeREP\footnote{http://wiserep.weizmann.ac.il/home} \citep{2012PASP..124..668Y}. 

The {\it Swift} satellite observed LSQ14an with the X-ray Telescope (XRT) on four epochs (PIs: Margutti \& Inserra)
listed in Table\,\ref{table:snx}.  For each XRT observation of \an\/ we extracted images over the 0.3-10 keV band and all data were reduced using the HEASARC\footnote{NASA High Energy Astrophysics Science Archive Research Center.} software package. 
  We measured the observed counts in each image in apertures of 5 pixel radius (11.8\arcsec), while the background in a large region (approximately 100 pixel radius), at a location free of bright X-ray sources. We then corrected this flux for the PSF contained within the aperture radius.  Swift also observed another nearby, slowly evolving SLSN, namely SN~2015bn. The monitoring of this was described in 
\cite{ni16a,ni16b} and analysis of the nebular spectra was presented alongside LSQ14an in \citet{je16}. For comparison, we retrieved the {\it Swift} X-ray data (5 epochs were taken, PI: Margutti) and applied similar reductions and measurements as for LSQ14an. 
We additionally combined all the available observations in a single stacked image in order to have a deeper measurement
All measurements for the two SLSNe are listed in Table~\ref{table:xr}.

\section{Host galaxy, location and redshift}\label{sec:host}

The host galaxy was detected in the Pan-STARRS1 3$\pi$ survey (it is outside the SDSS
footprint) in all five filters $grizy_{\rm P1}$ \citep[see][for a description of the filter set]{2012ApJ...750...99T}. 
Images in each of the filters were taken from the PS1 Processing
Version 2 (PS1.PV2) stacks. All the individual images which make up
these stacked images were taken
before MJD 56467 (2013 June 24). This is 4 months before the estimated
peak date of late October 2013 and hence there is unlikely to be
any SN flux in the images. Magnitude measurements of the host were carried
out using aperture photometry with an aperture of 2\arcsec, and the
zero points were determined with 15-18 PS1 reference stars 
\citep{2012ApJ...756..158S,2013ApJS..205...20M}
in the field resulting in the magnitudes listed in Table\,\ref{table:host}

As discussed in \cite{je16}, the position of LSQ14an was measured at
 $\alpha = 12^{\rm h}53^{\rm m}47^{\rm s}.82$, $\delta = -29^{\rm o}31'27''.6$ (J2000) 
in the astrometrically calibrated EFOSC2 images (with Sextractor). In comparison, the centroid of the host galaxy,
PSO J193.4492-29.5243, was measured at
 $\alpha = 12^{\rm h}53^{\rm m}47^{\rm s}.81$, $\delta = -29^{\rm o}31'27''.4$ (J2000). The difference between the two is
0.2\arcsec, which illustrates that LSQ14an is effectively coincident with the
dwarf galaxy host (as illustrated in Fig\,\ref{fig:host}). 
Ultraviolet (UV) photometry of the host was obtained with the {\it Swift} satellite + UltraViolet and Optical Telescope (UVOT) (programme ID 33205; P.I. Margutti). 
The  images  were co-added before aperture magnitudes were measured following the prescription of \citet{po08}.  A 5\arcsec\/ aperture was used to maximise the signal-to-noise ratio (S/N).  The UV magnitudes were determined by averaging the measurements of MJD 56740.52 (2014 July 03) and MJD 56998.41 (2014 December 07) when the SN flux was already dimmer than that of the host. We measured $uvw2 = 21.94 \pm0.14$, $uvm2 = 21.72 \pm0.15$, $uvw1 = 21.44 \pm0.16$ mag in the AB mag system. 

The host galaxy emission lines from LSQ14an are strong and conspicuously narrower than the SN emission features and they do not evolve in our spectral series. We used H$\alpha$ and other galaxy lines to refine the host redshift to 
$z=0.1637\pm0.0001$, equivalent to a luminosity distance of $D_{\rm L}=766$ Mpc assuming a fiducial cosmology of ${\rm H}_{0}=72$ \kms\/, ${\rm \Omega_{M}}=0.27$ and ${\rm \Omega_{\lambda}}=0.73$.
The foreground Galactic reddening toward \an\/ is E$_{\rm G}$(B-V) = 0.07 mag from the \cite{sf11} dust maps.
The available SN spectra do not show Na~{\sc id} lines from the host galaxy hence we adopt a total reddening along the line of sight towards \an\/ of E$_{\rm tot}$(B-V) = 0.07 mag.

To estimate physical diameters of the host galaxy we assumed (galaxy observed FWHM)$^{2}$ = (observed PSF FWHM)$^{2}$ + (intrinsic galaxy FWHM)$^{2}$. The measured FWHM of the galaxy and the PSF were 1.65\arcsec\/ and 
1.21\arcsec\/, respectively (with the stellar PSF FWHM averaged over 16 reference stars in {\it r}-band images, located within a 2\arcmin\/ radius of the host in the images). The angular size distance at this redshift ({\rm $D_{A}$} = 565.6 Mpc) 
implies a physical diameter of the galaxy of $\sim 3.09$ kpc.

\begin{figure}
\includegraphics[width=\columnwidth]{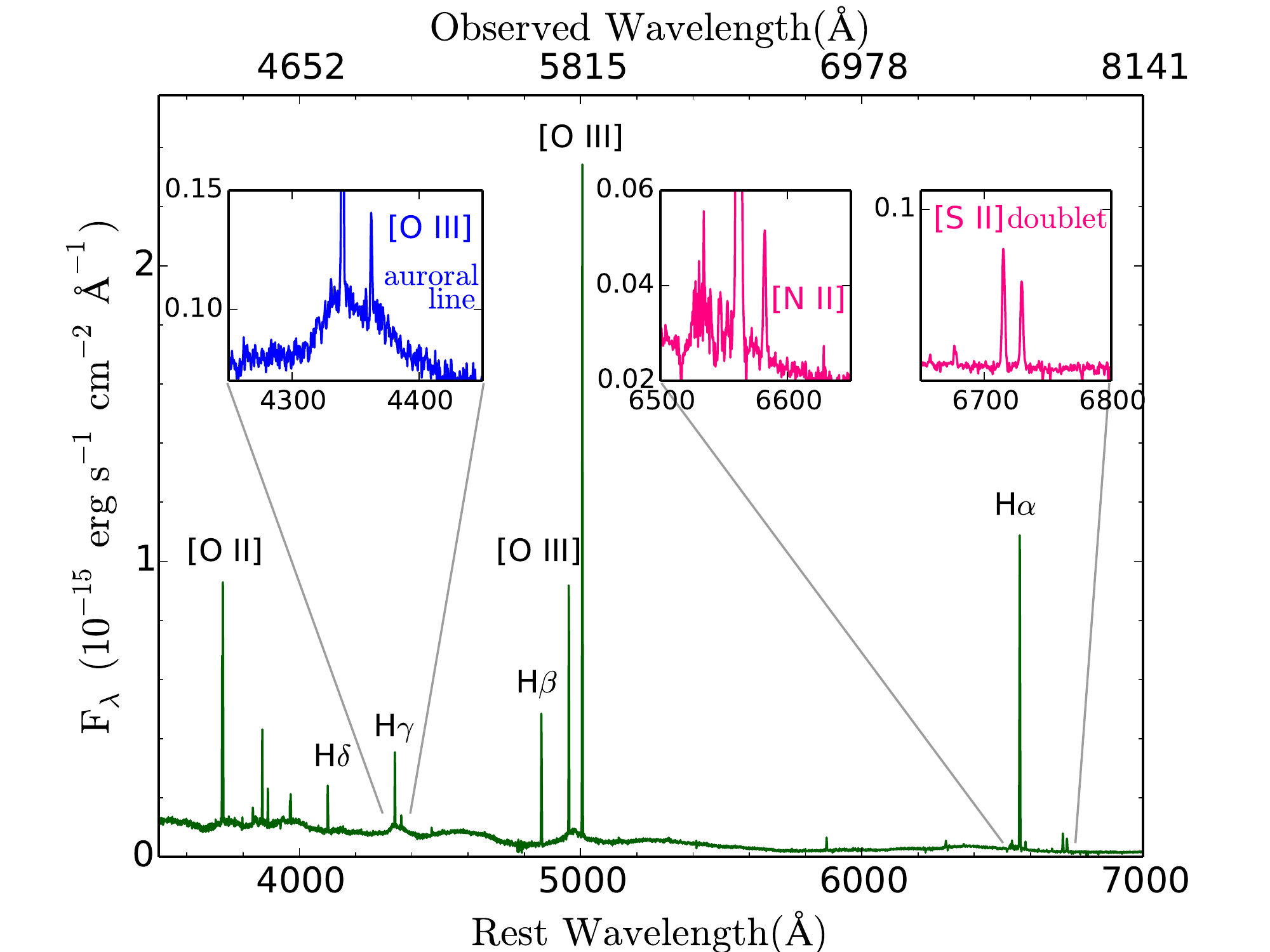}\\
\includegraphics[width=\columnwidth]{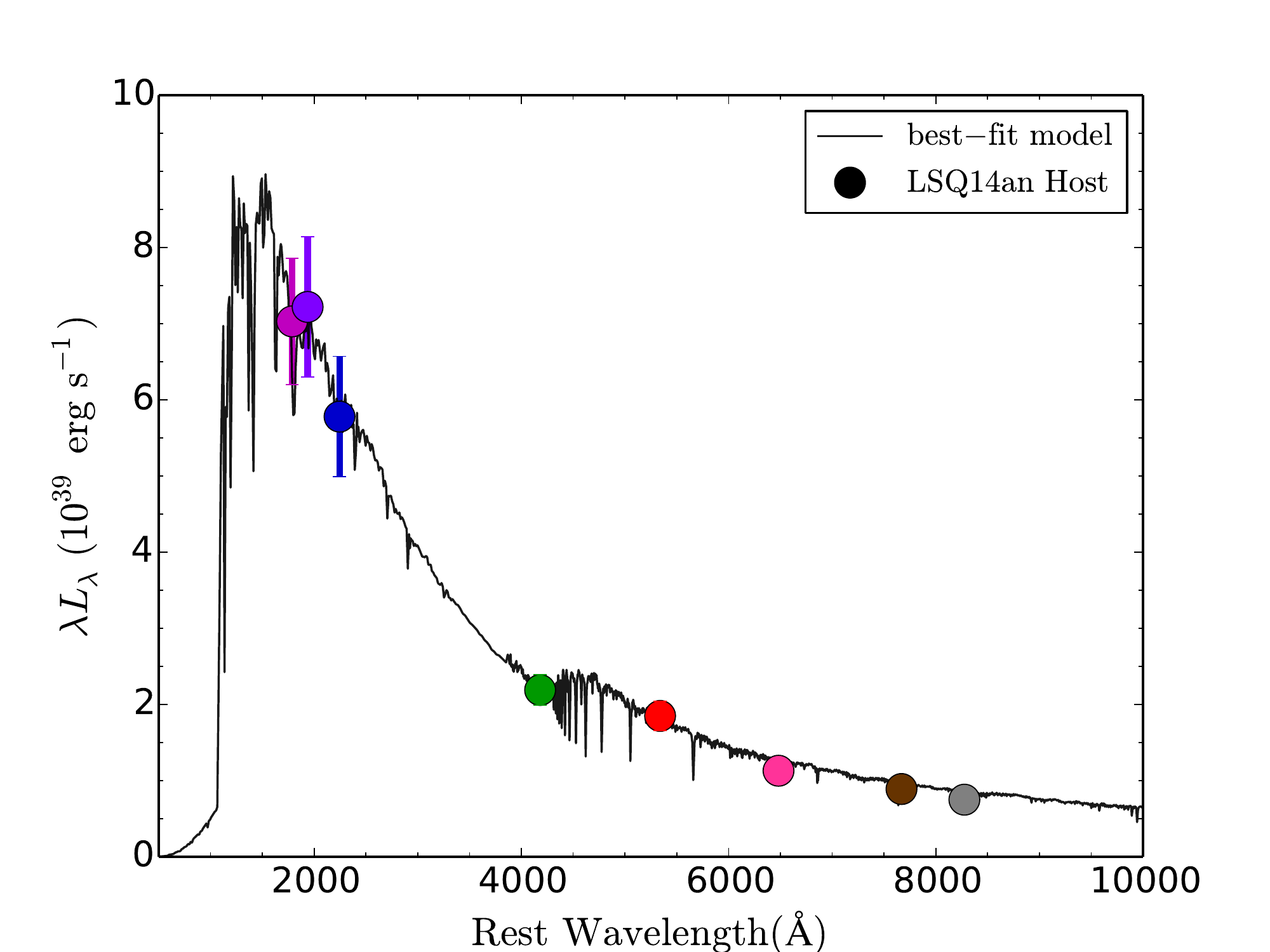}
\caption{{\it Top:} Spectrum of the host of LSQ14an taken by VLT+X-Shooter on 2014 June 21. The spectrum in the merged data from the UVB, VIS and NIR arms, corrected to rest-frame. Host lines are labelled in the spectrum. {\it Bottom:} The best-fit SED of the host of LSQ14an from {\sc magphys} \citep{2008MNRAS.388.1595D} shown in black line. The colour circles show photometry of LSQ14an host in the order of the Swift $uvw2, uvm2, uvw1$ and Pan-STARRS1 $grizy$ filters.}
\label{fig:hostsed}
\end{figure}

From the VLT+X-Shooter spectrum taken on 2014 June 21 (see Table~\ref{table:sp}) we measured the observed emission line fluxes of  PSO J193.4492-29.5243 (see Figure~\ref{fig:hostsed}, top panel).
Emission line measurements were made after fitting a low-order polynomial function to the continuum and subtracting off this contribution. This continuum flux is composed of the underlying stellar population of the host and the SN flux. No significant stellar absorption was seen in the spectrum, so we did not use stellar population model subtraction method. We fitted Gaussian line profiles of those emission lines using the QUB custom built {\sc procspec} environment within {\sc idl}. We let the full width at half-maximum (FWHM) vary for the strong lines and fixed the FWHM for weak lines adopting the FWHM of nearby stronger, single transitions. The equivalent width (EW) of lines were determined after spectrum normalisation. However, 
we caution that  the normalised continuum has  SN flux contamination and therefore the EW of each line is a minimum value
(rather than a real line strength compared to the galaxy continuum). 
Uncertainties were calculated from the line profile, EW and rms of the continuum, following the equation from \citet{1994ApJ...437..239G}. The observed flux measurement and related parameters are listed in Table\,\ref{tab:other_14an_flux_vlt}. The
 lines are identified as those commonly seen in star-forming galaxies, such as Green Peas \citep[e.g.][]{2012ApJ...749..185A}
and other SLSN hosts \citep[e.g. see][for a comprehensive line list for the host of PTF12dam]{ch15}.

We used the ``direct'' $T_{\rm e}$ method, following the prescription of \citet{2013ApJS..207...21N} and assuming Maxwell-Boltzmann distribution, based on the detection of \oiii\/ $\lambda4363$ line and found $T_{\rm e}$(\oiii)$=$13700$\pm$1200 K.
We can then infer a $T$[O~{\sc ii}] = 13100$\pm$800 K, resulting in an oxygen abundance of 12 + $\log$(O/H) = $8.04 \pm 0.08$,  equivalent to 0.22 Z$_{\odot}$ considering a solar value of 8.69 \citep{2009ARA&A..47..481A}. 
We also checked the electron temperature at the lower-ionisation zone using the ratio of the nebular lines \oii\/ $\lambda\lambda7320, 7330$ and thus retrieved $T_{\rm e}$(\oii) of $\sim$7100~K which is much lower than the inferred $T$[O~{\sc ii}]. We believe the discrepancy is due to the low S/N of the \oii\/ lines with a large SN flux contamination and hence we do not use the $T$[O~{\sc ii}].
We used the open-source {\sc Python} code {\sc pymcz} from \citet{bianco16} to alternatively calculate metallicity of the host of LSQ14an with different strong line calibrations, and we listed them as follows:
12 + $\log$(O/H) = $8.02 \pm 0.02$ from the N2 method \citep[][]{pp04} and 12 + $\log$(O/H) = $7.96 \pm 0.01$ of the O3N2 method \citep[][]{pp04}. The R23 value of 0.9 is at an insensitive regime, so we do not use this scale here.

We employed the {\sc magphys} stellar population model program of \citet{2008MNRAS.388.1595D} to estimate the stellar mass from the observed photometry (after foreground extinction correction) of the host galaxy.
{\sc magphys} gives the best-fit model spectrum ($\chi^{2}_{\rm red} = 0.316$) and the total stellar mass of the hosts (see Figure~\ref{fig:hostsed}). The median and the $1\sigma$ range of the stellar mass of \an\/ host are  
$4.0^{+2.5}_{-1.1} \times 10^{8}$ \M\/, which is consistent with the measurements of \citet{sch16} for the same host, and that retrieved in SLSNe hosts \citep[e.g.][]{lu14,le15,pe16}.

We converted the \Ha\/ luminosity of $2.41 \times 10^{41}$ erg s$^{-1}$ to calculate the star-formation rate (SFR) using the conversion of \citet{1998ARA&A..36..189K} and a Chabrier IMF\footnote{The tool employed to evaluate the SFR uses a Chabier IMF, while \citet{1998ARA&A..36..189K} a Salpeter IMF. We took that in account dividing the SFR by a factor of 1.6. }.
We found a SFR of LSQ14an host SFR = 1.19 \M\/ yr$^{-1}$, which is higher than the average, but still consistent with other slow-evolving SLSNe I \citep[e.g.][]{ch15,le15}. In this case our estimate is somewhat lower than that reported by \citet{sch16}.
Considering the stellar mass previously evaluated, we determined the specific SFR (sSFR) of 2.98 Gyr$^{-1}$ for the host of LSQ14an. A summary of \an\/ host galaxy properties is reported in Table~\ref{table:host}.

\begin{table}
\centering
\caption{\an\/ host galaxy observed magnitudes (AB system) and properties derived from the spectral energy distribution and late time spectra (see Appendix~\ref{sec:host} for further information).}
\begin{tabular}[]{llll}
\hline
$uvw2$ (mag)& $ 21.94 \pm 0.14$\\
$uvm2$ (mag)& $  21.72 \pm 0.15$\\
$uvw1$(mag)& $  21.44 \pm 0.16$ \\
$g_{\rm P1}$(mag)& $  20.93\pm  0.11$\\
$r_{\rm P1}$(mag)& $ 20.50\pm   0.12$\\
$i_{\rm P1}$(mag)& $ 20.56\pm   0.10$\\
$z_{\rm P1}$(mag) & $ 20.42\pm   0.14$\\
$y_{\rm P1}$(mag)& $ 20.43 \pm 0.19$\\  
\hline
Physical diameter (kpc) & 3.09  \\ 
H$\alpha$ Luminosity (erg s$^{-1}$) &$2.41 \times 10^{41}$ \\
SFR ($M_{\odot}$ yr$^{-1}$) &1.19\\
Stellar mass (log(M/$M_{\odot}$)) &  8.6$^{+0.2}_{-0.1}$ \\
sSFR (Gyr$^{-1}$) & 2.98 \\
$12+\log {\rm(O/H)}$ ($T_{\rm e}$) & $8.04 \pm 0.08$ \\
$12+\log {\rm(O/H)}$ (PP04 N2) & $8.02 \pm 0.03$ \\
$12+\log {\rm(O/H)}$ (PP04 O3N2) & $7.96 \pm 0.01$\\
$\log$ (S/O) & $-1.94 \pm 0.15$ \\
$\log$ (N/O) ($T_{\rm e}$) & $-1.74 \pm 0.18$ \\ 
\hline 
\end{tabular}
\label{table:host}
\end{table}

\begin{figure}
\includegraphics[width=\columnwidth]{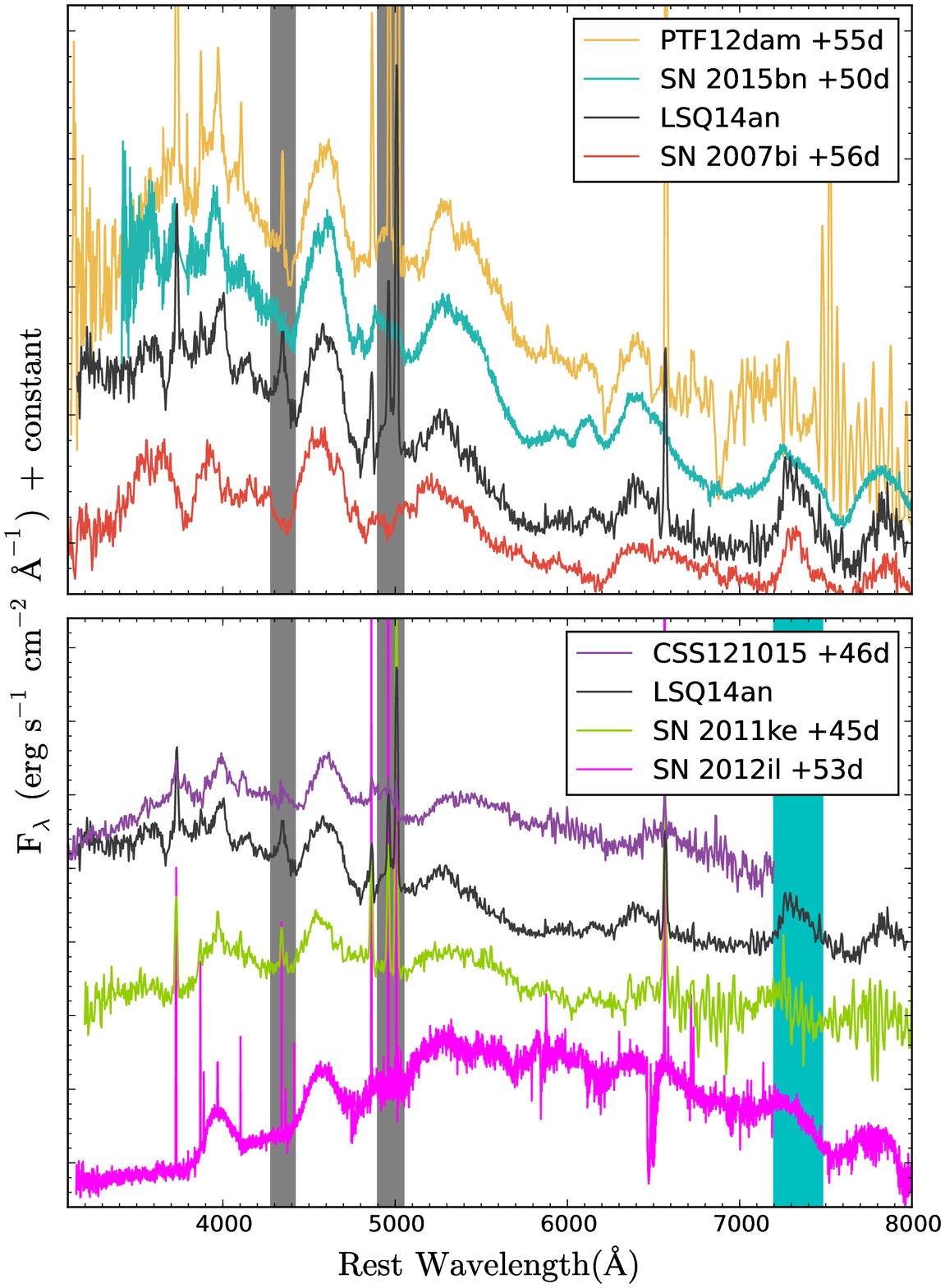}
\caption{Top: comparison of \an\/ classification spectrum with those of other slow-evolving SLSNe I \citep[SN 2007bi, PTF12dam, SN 2015bn;][]{gy09,yo10,ni13,ni16a} around 50 days post maximum suggesting \an\/ was discovered at $\sim$55 days from maximum light. Bottom: same \an\/ spectrum of the top panel compared with fast-evolving SLSNe I  \citep[SN~2011ke, SN~2012il and CSS121015;][]{in13,ben14} in order to highlight the differences. The grey area identify the \oiii\/ lines observed in emission in \an\/ spectra, while the cyan in the bottom panel identifies the region of the [Ca~{\sc ii}] $\lambda\lambda$7291, 7323.}
\label{fig:spcmp}
\end{figure}

\begin{figure*}
\includegraphics[width=18cm]{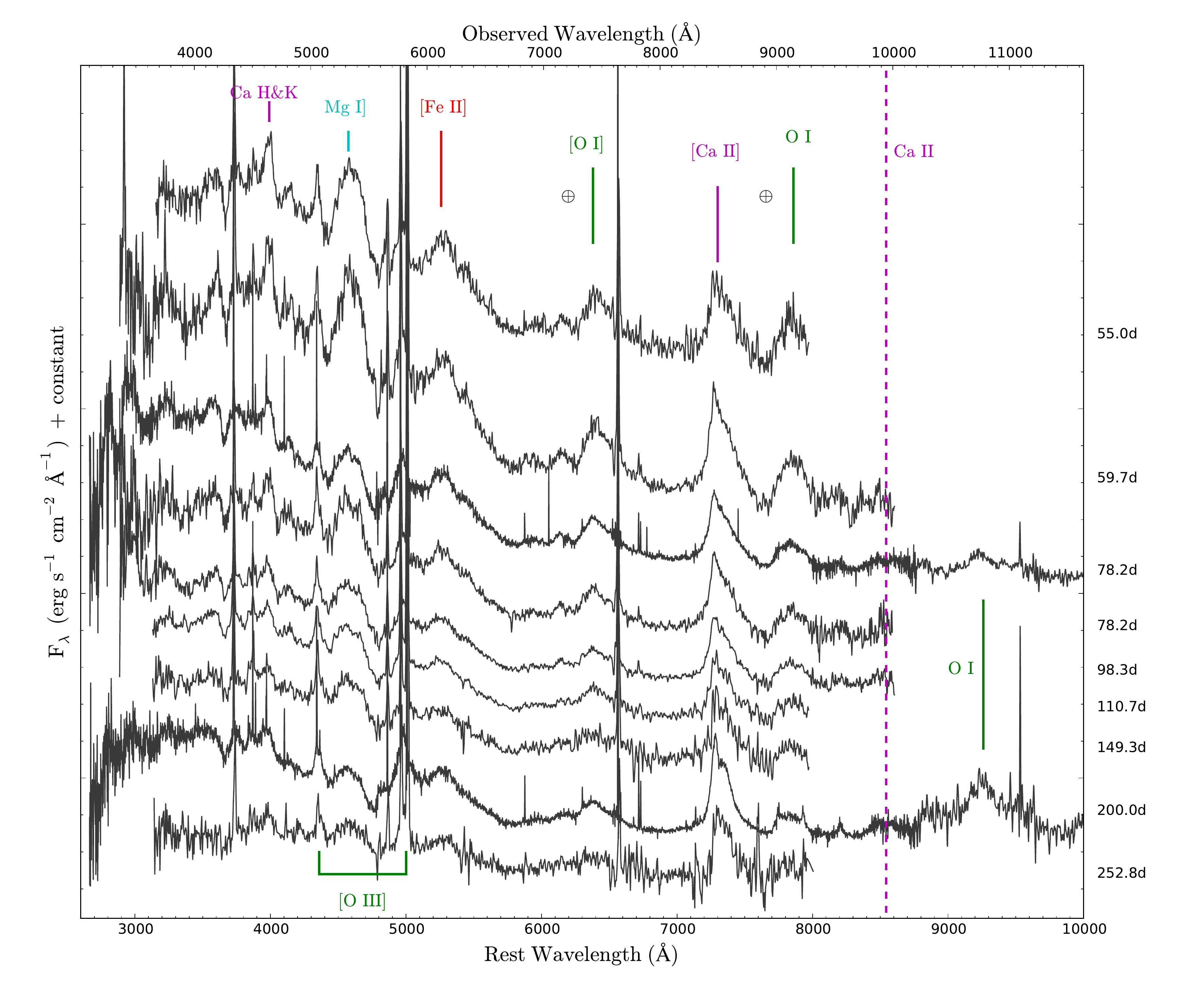}
\caption{Spectral evolution of \an. The phase of each spectra is relative to the assumed peak epoch. The $\oplus$ symbol marks the positions of the strongest telluric absorptions. The most prominent features are labelled, together with the position (dashed magenta line) of the NIR Ca~{\sc ii}, which is not observed.}
\label{fig:spev}
\end{figure*}

\section{Spectra}\label{sec:sp}
The first spectra taken of \an\/ by PESSTO \citep{atel} already suggested an 
epoch well after maximum brightness.  The absence of the characteristic broad  O~{\sc ii}
absorption lines observed in SLSNe, at and before the peak epoch, together with a temperature of $\sim$8000~K - derived from the blackbody fit to the continuum of our spectra - confirm the initial classification phase. In order to secure the phase with more precision and estimate the date of the peak of the light curve, we compared our classification spectrum with other slow-evolving SLSNe I. As highlighted in the top panel of Figure~\ref{fig:spcmp}, the  \an\/ spectrum taken on the 2nd January 2014 is very similar to those of the three other objects with broad light curves: SN~2007bi, PTF12dam and SN~2015bn at 50 to 60d after peak. 

At this phase, the emission lines that are developing and becoming pronounced are 
similar to those that have quantitatively been identified and modelled by \cite{je16}. 
The Mg~{\sc i}] $\lambda$4571 line is strong, and this usually begins to appear about $\sim$35 days post peak \citep[][]{in13, ni16a}. 
The region between 5000~\AA\/ and 5800~\AA\/ shows a broad emission feature
which is likely a blend of
Mg~{\sc i} $\lambda$5180, 
[Fe~{\sc ii}] $\lambda$5250 and 
[O~{\sc i}] $\lambda$5577.  
There is a distinct lack of absorption that could be attributed to 
Si~{\sc ii} $\lambda$6355 and the forbidden doublet of [O~{\sc i}] $\lambda\lambda$6300,~6363 is beginning to emerge in emission (see  Figure~\ref{fig:spev} for line identification).  Finally, the [Ca~{\sc ii}] $\lambda\lambda$7291,~7323 line strength and the equivalent width of O~{\sc i} $\lambda$7774 both support an epoch of the spectrum of 
 +50-55 days after maximum in the rest-frame\footnote{All the phases are reported in rest-frame unless otherwise stated.}. As a consequence, assuming a phase of +55d for the first spectrum, we estimate the peak epoch to have been MJD $56596.3\pm5.0$  (around 2013 Oct 31). This date is consistent with 
our analysis of  the light curve evolution (see Section~\ref{sec:lc}).

For comparison, we also showed the earliest PESSTO spectrum of LSQ14an together with 
faster evolving SLSNe (bottom panel of Figure~\ref{fig:spcmp}). 
SN~2011ke and SN~2012il decline much faster than the other three comparison objects and are similar to 
SN~2010gx in their overall evolution \citep[see][for a discussion of these two objects]{in13}.  CSS121015 is a peculiar SLSN I which 
shows weak signs of interaction with a hydrogen CSM that fade with time, displaying broad lines similar to those of SLSNe I  \citep{ben14}. In this comparison, the 
noticeable differences are that the Mg~{\sc i}] line has a larger equivalent width 
(a factor of 2 higher in LSQ14an than the others), and the fast-evolving SLSNe do not show 
the forbidden [O~{\sc i}] and [Ca~{\sc ii}] at this phase. In truth, a clear detection of [Ca~{\sc ii}] has only been recently reported for a fast-evolving SLSNe \citep[at +151d from maximum in Gaia16apd,][a SLSNe that shows a late photometric behaviour intermediate between the fast- and slow-evolving events]{kan16}, but see Section~\ref{sec:ca2} for a more detailed analysis of the elements contributing to the feature. The continuum of 
LSQ14an, and the other slowly evolving objects above are also significantly bluer than the fast-evolving and stay blue for longer (more than a factor of 3 in time) as already noted for SN~2015bn by \citep{ni16a}. The prominent emission feature at  5200~\AA\/ to 5600~\AA\/ has a 
markedly different profile shape in LSQ14an - sharper with a peak consistent with it being a 
blend of Mg~{\sc i} $\lambda$5180 and [Fe~{\sc ii}] $\lambda$5250. This profile is similar to that marking the beginning of the pseudo-continuum, dominated by Fe lines, in interacting SNe such as SNe 2005gl and 2012ca \citep{gy07,in16a} but in the SLSNe case the feature is 200\AA\/ bluer and not as sharp as in the interacting transients.

\begin{figure}
\includegraphics[width=\columnwidth]{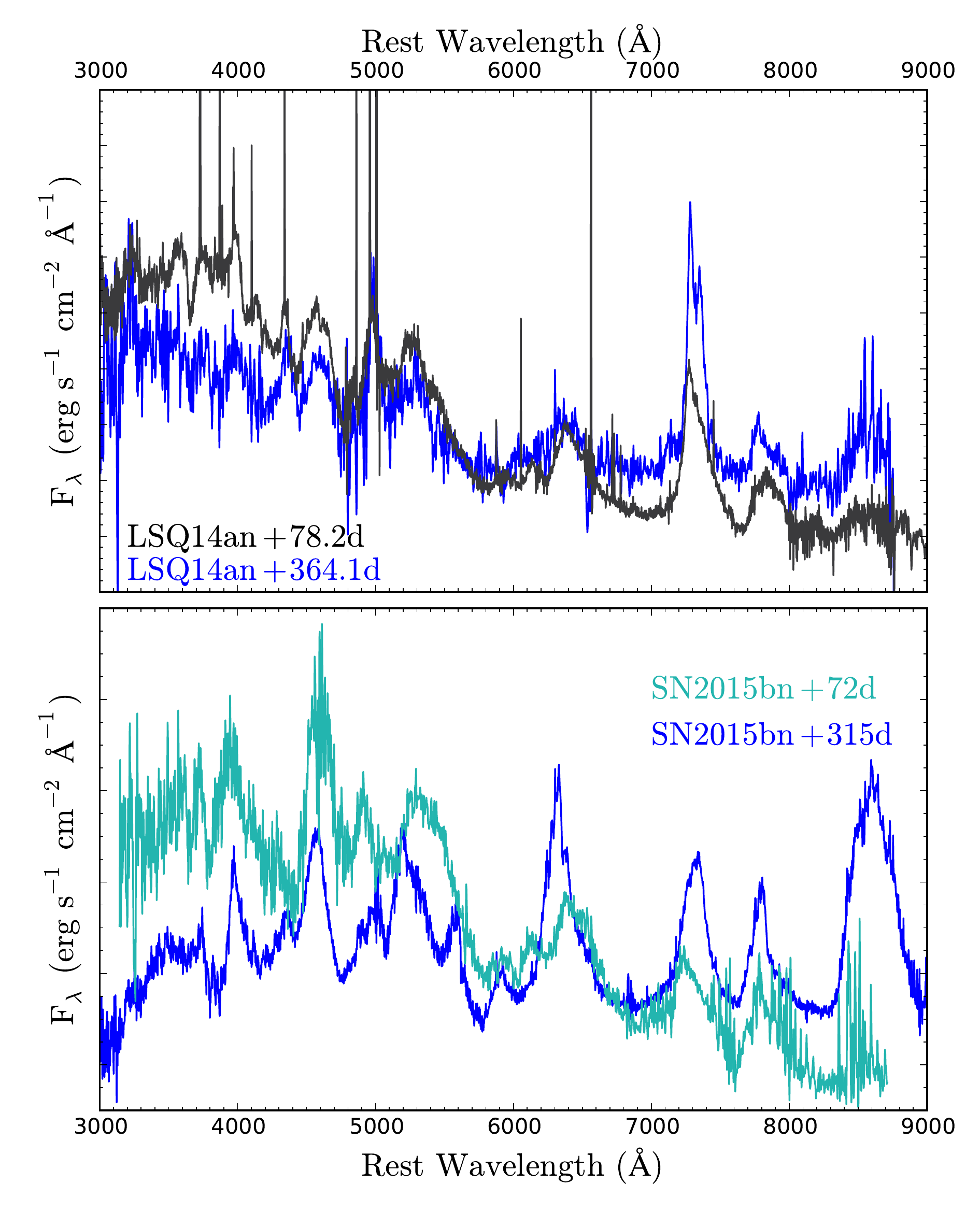}
\caption{{\it Top:} \an\/ at +78d (black) and +364d (blue, after subtraction of the galaxy model,and removal of host galaxy
lines). Note that the +364d spectrum \citep{je16} has been scaled by 2.3 to match the earlier one. The lines have not changed in almost a year. The same is true if the +50d and +409d \citep{je16} spectra are compared. {\it Bottom:} similar comparison of before but for SN~2015bn \citep{ni16a,je16}.}
\label{fig:cmpel}
\end{figure}

In Figure~\ref{fig:spev} we show the optical spectroscopic evolution of \an\/  with nine spectra
taken from  +55d to +253d after the estimated peak epoch. At all epochs, the spectra are ``pseudo-nebular'', showing permitted, semi-forbidden and forbidden lines.  However it is 
remarkable that the lines that are unambiguously
identified in \cite{je16} through modelling the 
nebular phase spectra at +300-400d are
already visible and prominent at +50d (cfr. Figure~\ref{fig:cmpel}). 
We note that the epoch of the first spectrum is 
supported by the previous spectroscopic comparison of Figure~\ref{fig:spcmp} and also by \an\/ photometric evolution (cfr. Section~\ref{sec:lc}).
The strongest emission line features are 
almost all the same in these spectra which are
separated by a long time interval up to a year 
\citep[see Figure~\ref{fig:cmpel} in this paper and fig.\,7 of][]{je16}. The strong blue 
continuum is a feature of both LSQ14an, and 
SN~2015bn even after subtraction of the blue
host galaxy. A similar behaviour is also shown by SN~2015bn on an almost equivalent time baseline of \an\/ (Figure~\ref{fig:cmpel}). The strongest emission lines identified at late time are already in the +72d spectrum \citep[and in any spectrum after $\sim$50d post peak; see][]{ni16a}. 
The only small difference is the [O~{\sc i}] $\lambda\lambda$6300, 6364 that appears later in SN~2015bn \citep{je16}. We note that the continuum flux of the host galaxy becomes comparable to the SN at +150d and 
later, therefore the last two spectra at +200d and +252d will contain host galaxy flux. 

The near infrared coverage of X-Shooter 
at +78d and +200d allows the 
identification of the line at 9200~\AA\/ line as O~{\sc i} $\lambda$9263. 
This line was identified as
O~{\sc i} 
 in the modelling of the
+315d spectrum of SN~2015bn and the +365d 
spectrum of LSQ14an by  \cite{je16}. This
is a recombination line 
 (2s$^2$2p$^3$($^4$S$^{\rm o}$)3p) 
and we measure a  
full width at half maximum (FWHM) of 
$v\sim8000$ \kms\/ in the +200d spectrum. The O~{\sc i} $\lambda$7774 is another recombination line which decays from the lower state of the  O~{\sc i} $\lambda$9263 transition. 
As expected, the two lines have the same velocity. 

We also measured the FWHM velocity of the  [O~{\sc i}] and [Ca~{\sc ii}], which are usually 
the strongest forbidden lines in nebular spectra of supernovae. They display an almost constant velocity with average $v\sim7500$ \kms\/ for [O~{\sc i}] and  $v\sim7400$ \kms\/ for [Ca~{\sc ii}].
The calcium line, clearly not Gaussian, is noticeably asymmetric (see Section~\ref{sec:ca2}) throughout the evolution and is much stronger than [O~{\sc i}]. 
The feature could be a blend of [O~{\sc ii}] $\lambda\lambda$7320, 7330
and  [Ca~{\sc ii}] $\lambda$7291,7323. A more detailed discussion of this and the implications 
for the explosion mechanism are presented in \citet{je16}.  However, we already notice that lines coming from the same elements show different velocities and how semi-forbidden and forbidden lines are observed earlier and at higher temperatures than normal stripped envelope SNe. 

What is surprising is the absence of the Ca~{\sc ii} NIR triplet $\lambda\lambda 8498, 8542, 8662$ up to 200 days (see Figure~\ref{fig:spev}), whereas it is present at +365d in \citet{je16}. Since it requires high temperatures, this feature is usually seen quite early and disappears after 100 days. On the contrary, in \an\/ the evolution is the opposite and in some extent similar to that of SN~2015bn where the NIR Ca grows in strength from 100 days to more than 300d \citep{ni16b,je16}. 
The two X-Shooter spectra 
have a coverage out to 2.0$\mu$m in the rest-frame, but they are featureless beyond 
1.0$\mu$m and display no strong and identifiable emission lines. 

The narrowest emission lines which are lines observable throughout the spectroscopic evolution, are the Balmer, \oii\/ and  \oiii\/ lines  due to the host galaxy. They have a constant velocity of $v=150\pm20$ \kms\/ (measured from the two X-Shooter spectra) comparable with those of
the host galaxies of SLSNe in previous studies \citep[e.g.][]{le15}.

\begin{table}
\caption{Breakdown of the oxygen and calcium lines analysis from our X-Shooter spectra of \an\/ (see Sections~\ref{sec:o3}~\&~\ref{sec:ca2}).}
\begin{center}
\begin{tabular}{lcc}
\hline
Phase (day) & +78.2 & +200.0 \\
\hline
& [O~{\sc i}]~$\lambda\lambda$6300,~6363 & \\
shift (\kms) & - & - \\
$v$ (FWHM, \kms) & 7500 & 7500\\
$n_{\rm e}$ (cm$^{-3}$) $^{\dagger}$ & $\sim$10$^9$& $\sim$10$^9$ \\
\hline
 & \oiii\/ $\lambda4363$ &\\
shift (\kms)$^*$ & -1400 & -1400 \\
$v$ (FWHM, \kms) & 3200 & 3200\\
$n_{\rm e}$ (cm$^{-3}$)  & $\sim$10$^7$& $\sim$10$^7$ \\
\hline
& \oiii\/~$\lambda\lambda$4959,~5007 & \\
shift (\kms)$^*$ & -1400  & -1400 \\
$v$ (FWHM, \kms) & 3500 & 3500 \\
$n_{\rm e}$ (cm$^{-3}$)  & $\sim$10$^7$& $\sim$10$^7$ \\
\hline
& [Ca~{\sc ii}] $\lambda\lambda$7291, 7323& \\
shift (\kms)$^*$ & -1300& 0 \\
$v$ (FWHM, \kms) & 7400 & 7400 \\
$n_{\rm e}$ (cm$^{-3}$)$^{\dagger}$& $>$10$^9$ & $>$10$^9$ \\    
\hline
& [O~{\sc ii}] $\lambda\lambda$7320, 7330 +[Ca~{\sc ii}] $^{\ddagger}$& \\
shift (\kms)$^*$ & -1900/-1700& -1400/0 \\
$v$ (FWHM, \kms) & 1700/2800& 1700/2800\\
$n_{\rm e}$ (cm$^{-3}$)$^{\dagger}$& $\sim$10$^9$/$>$10$^9$ & $\sim$10$^9$/$>$10$^9$\\                                                                                                                                
\hline
\end{tabular}
\end{center}
* w.r.t. the centroid of the line or main doublet\\
$\dagger$ data derived from the estimates reported in \citet{je16}\\
$\ddagger$ left values are related to [O~{\sc ii}], while the right to [Ca~{\sc ii}] 
\label{table:lines}
\end{table}%

\subsection{[O~{\sc iii}] $\lambda$4363 and $\lambda\lambda$4959, 5007 lines}\label{sec:o3}
\begin{figure}
\includegraphics[width=\columnwidth]{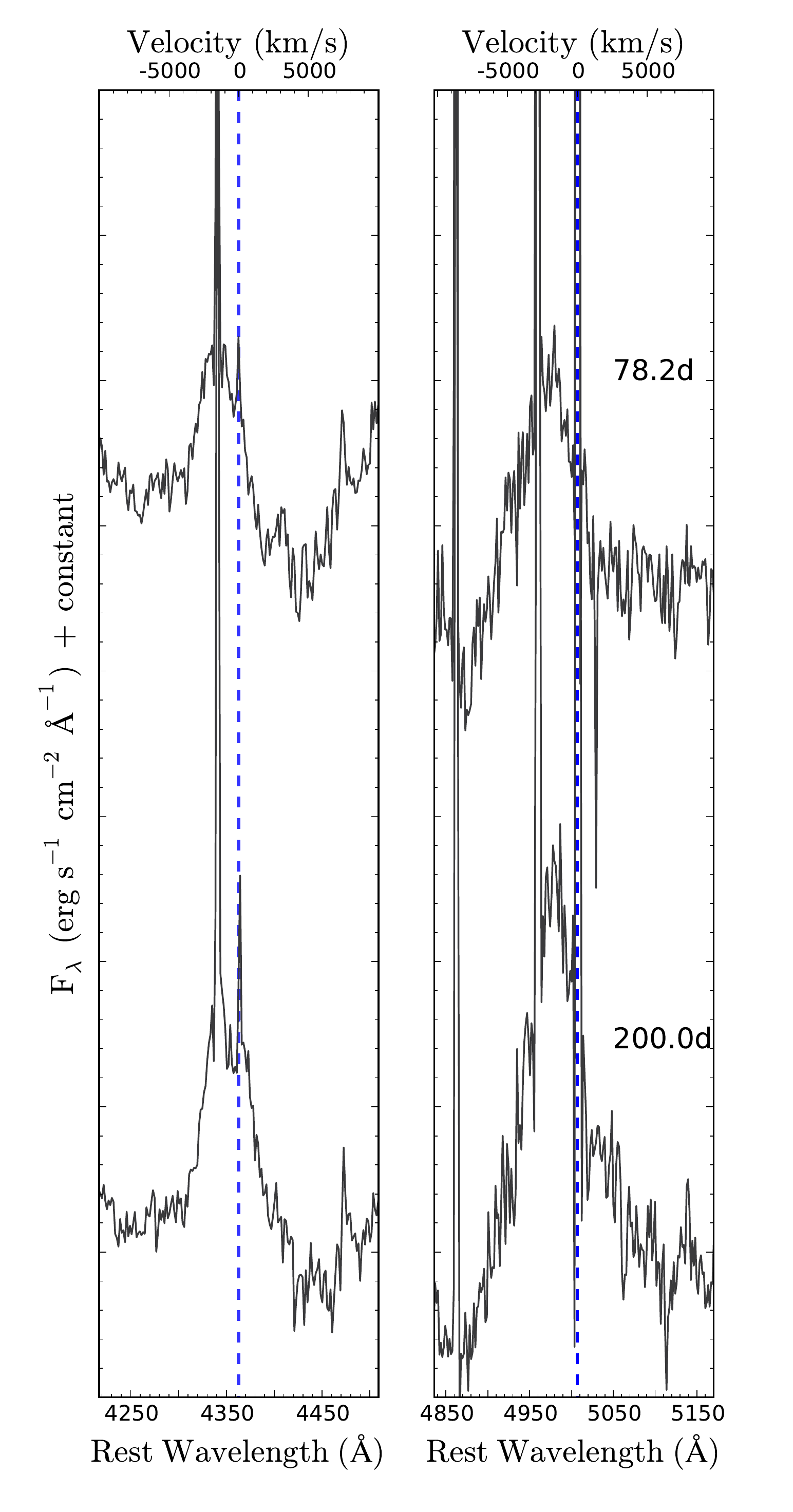}
\caption{{\it Left}: \an\/ [O~{\sc iii}] $\lambda$4363 evolution from the two X-Shooter spectra at +78.2 and + 200.0 days. {\it Right}: \an\/ [O~{\sc iii}] $\lambda\lambda$4950,~5007 evolution. The zero velocity is set at 5007~\AA\/, since that is the strongest component of the [O~{\sc iii}] doublet. The vertical dashed blue lines denote the centroids of the lines. H$\gamma$ (right panel) and H$\delta$ (left panel) are also visible.}
\label{fig:ox}
\end{figure}

A peculiarity shown by the spectroscopic evolution of \an\/  
is the presence of a broad component of \oiii\/ $\lambda$4363 and $\lambda\lambda$4959, 5007 (see Figure~\ref{fig:ox}).
In the first two 
PESSTO spectra with EFOSC2, these oxygen lines are prominent but the resolution of
the Grism\#13  prevented easy identification of emission broader than the host
galaxy lines (see Figure~\ref{fig:ox}). The first X-Shooter spectrum taken at +78d  shows that 
\oiii\/ $\lambda$4363 and $\lambda\lambda$4959, 5007 have a broad component 
arising in the supernova ejecta. The existence and origin of these lines as 
broad \oiii\/ was  first pointed out by \citet[][for PS1-14bj which is a very slow-evolving SLSN]{lu16} and they have been analysed in 
detail in \cite{je16} in the later nebula spectra. 

The average velocities of the components, as measured from the two X-Shooter spectra  are $v_{\rm \lambda4363}\sim3200$ \kms\/ and $v_{\rm \lambda\lambda4959,5007}\sim3500$ \kms\/ (see also Figure~\ref{fig:ox}).
Although these broad lines are severely blended with the galaxy emission lines in the lower
resolution PESSTO spectra, we measure no velocity evolution in the FWHM of the features
over the whole period (similarly to the other lines discussed in Section~\ref{sec:sp}). 
The velocities of \oiii\/ 
we measure from the X-Shooter spectra are a factor 2-3 less than those of the [O~{\sc i}] forbidden lines and  the O~{\sc i} recombination  lines. Even if we assume a low-mass, dilute oxygen region, allowing a higher energy deposition than the other slow-evolving SLSNe \citep{je16}, this would not take in account their intrinsic different velocities, which could be explained with multiple emitting regions. Something similar was suggested in case of SN~2015bn but for different oxygen lines \citep{ni16b}.

The measured ratio for these collisionally excited lines in the optically thin limit can provide some information about the electron density of the emitting region where those are formed 

\begin{equation}
\frac{f_{\rm \lambda\lambda4959,5007}}{f_{\rm \lambda4363}} = \frac{7.90\, {\rm exp}(3.29 \times 10^{4}/T)}{1+(4.50 \times 10^{-4} \times n_{\rm e}/T^\frac{1}{2})}\, ,
\end{equation}

from \citet{of}, where $T$ is the temperature and $n_{\rm e}$ is the electron density of the medium. We measured 
$f_{ \lambda\lambda4959,5007}/f_{ \lambda4363} =1.8$ from our X-Shooter spectra on an average of two measurements with a Gaussian fit. The ratio is consistent with that reported by \citet{lu16} for PS1-14bj ($<3$) and hence less than what is typically observed in gaseous nebulae \citep{of} and nebular type IIn \citep[e.g. SN~1995N,][]{fr02}. 
We initially assumed the ejecta temperature ($T\sim$8000 K) and found $n_{\rm e}=6.2\times 10^7$ cm$^{-3}$ for the emitting region of the \oiii\/ lines.
 If we increase the assumed temperature to $T=20000~K$ the density decreases to $n_{\rm e}=8.0\times 10^6$ cm$^{-3}$. 
By comparison, 
the [O~{\sc i}] $\lambda\lambda$6300, 6363 arising from the inner region of the ejecta has $n_{\rm e}\sim10^9$ cm$^{-3}$ \citep[as measured for late spectra of \an\/, $\sim$200d later, by][and having taken in account that density evolves as $t^{-3}$]{je16}. These lines are mainly collisionally excited with thermal ions 
at all times, as they are close to the ground state \citep[see][for a more in depth analysis]{je16}.  

Although [O~{\sc iii}] and [O~{\sc i}] can come from the same physical region, the different densities ($n_{\rm e}$(\oiii)$<$$n_{\rm e}$([O~{\sc i}]), velocities ($v$(\oiii)$<$$v$([O~{\sc i}])) and the absence of a comparable strong [O~{\sc ii}] $\lambda$7300 would suggest that these lines are formed in two different regions (see Section~\ref{sec:dis} for the interpretation).

In the magnetar scenario, the \oiii\/ lines could come from an oxygen-rich ejecta layer which is ionised from the X-ray energy coming from the inner engine. This would be very similar to the scenario of a pulsar wind nebula expanding in a H- and He-free supernova gas, for which the second strongest lines predicted are \oiii\/ $\lambda$4959, 5007 \citep{ch92}. On the other hand, the strongest lines observable in a H-free pulsar wind nebula should be those of [S~{\sc iii}] $\lambda\lambda$9069, 9532 \footnote{We note that if hydrogen is also present in the nebula, \oiii\/ $\lambda$5007 and [S~{\sc ii}] $\lambda$6723 are stronger than [S~{\sc iii}] \citep{ki89}. However, in our case also [S~{\sc ii}] is not observed.}. 
These are not observed in our NIR spectra, suggesting it may not  be the X-ray heating from the magnetar that causes the presence of the \oiii\/ lines. 

As shown in Figure~\ref{fig:ox}, we also note that the peak of the \oiii\/ $\lambda$4363 is blueshifted by $\sim1400$ \kms\/ 
compared to the rest-frame of the SN and the centroid of the \oi\/ lines (see Table~\ref{table:lines} for a comparison between the velocities of the oxygen lines). Similarly the 
 \oiii\/  $\lambda\lambda$4959,~5007 lines have a centroid which appears shifted of $\sim1400$ \kms\/. 
These lines and shifts are clear in the high resolution X-Shooter spectra, but are not easily discerned in the lower resolution EFOSC2 spectra. 

\begin{figure}
\includegraphics[width=\columnwidth]{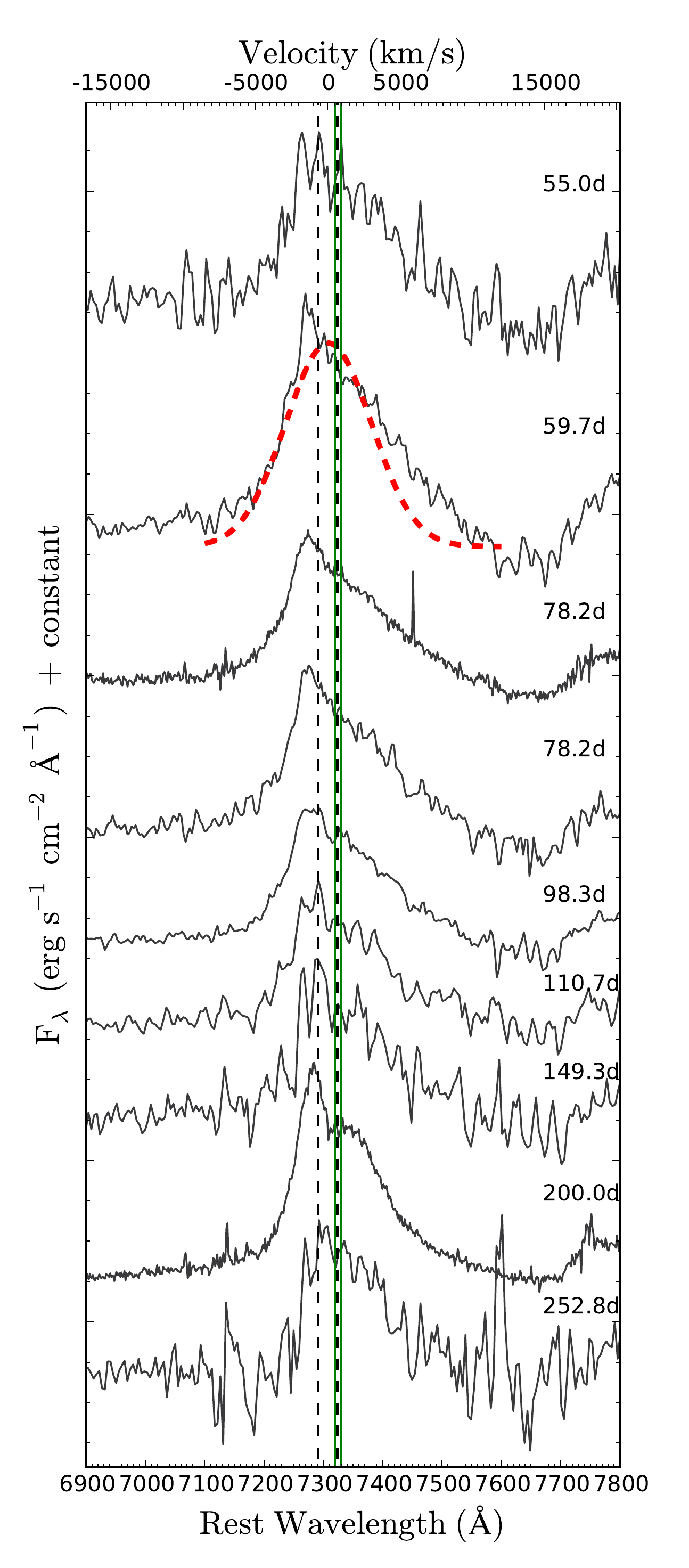}
\caption{\an\/ [Ca~{\sc ii}]+[O~{\sc ii}] evolution from the first epoch at +55.0 days to the last at +252.8d. A double gaussian fit (red dashed line) to [Ca~{\sc ii}] is shown for the +59.7d epoch to highlight the skewed shape of the profile. The two vertical black dashed lines denote $\lambda\lambda$7291, 7323 at rest-frame, while the solid green mark [O~{\sc ii}] $\lambda\lambda$7320, 7330. The zero velocity is set at the centroid of the [Ca~{\sc ii}] doublet. Phase with respect to the maximum is also shown.}
\label{fig:cal}
\end{figure}

\subsection{Forbidden lines around 7300\AA}\label{sec:ca2}
Explosively produced calcium would likely come from an O-burning zone rather than a C-burning zone, where the 
[O~{\sc i}] $\lambda\lambda$6300, 6363 lines would form. 
In slow-evolving SLSNe [Ca~{\sc ii}] $\lambda\lambda$7291, 7323
 has been observed in emission from surprisingly early stage:  about 50-70 days from maximum \citep[e.g. SNe 2007bi, 2015bn;][]{gy09,yo10,ni16a}. We see similar behaviour from  \an. [Ca~{\sc ii}] appears almost at the same phase in normal stripped envelope SNe \citep[after 60-70 days, e.g. SNe 1994I, 2007gr;][]{fi95,hun09} and later in broad line SNe \citep[after 100d, e.g. SN 1998bw;][]{pat01}, which however have less massive ejecta than the slow-evolving SLSNe. Also in fast-evolving SLSNe it appears at least after 100 days from peak \citep{kan16}. 
Hence in SLSNe, where it is detected, it seems to appear earlier than expected. 

The  [Ca~{\sc ii}] profile shows a component which has a peak at a velocity of $-1300$ \kms\/
with respect to rest-frame [Ca~{\sc ii}]. Although we are somewhat limited by the EFOSC2 resolution, the 
blue shifted peak is visible in all spectra from +55d to +253d (see Figure~\ref{fig:cal}). 
The X-Shooter spectra show the line profile well resolved and indicate that this is not a
simple blue shifting of the peak of [Ca~{\sc ii}]. The three most likely explanations are 
either asymmetry in the [Ca~{\sc ii}] emitting region, a second [Ca~{\sc ii}] emitting region which is at a different velocity, 
or another ionic component.  
Figure~\ref{fig:cal} shows a double gaussian fit to the 
broad [Ca~{\sc ii}] in the second epoch (FWHM= 7400 \kms\/) with the blue peak visible. 
If this is [Ca~{\sc ii}], then the blueshift of $-1300$~\kms\/ is similar to that observed in the \oiii\/ lines. 
Therefore the region that is producing the  \oiii\/ lines may also be emitting [Ca~{\sc ii}]. 

There is an iron line, [Fe~{\sc ii}] $\lambda$7155, but this transition is 120\AA\/ too far blue and 
the X-Shooter profile would not support that identification. Furthermore we would also have expected to observe [Fe~{\sc ii}] $\lambda$1.257 $\mu$m with a similar strength, but the spectra at that wavelength are featureless. 
\cite{je16} showed that the strength of the overall emission feature of [Ca~{\sc ii}]  is much stronger in LSQ14an than in the other two comparator objects (SN~2015bn and SN~2007bi). Their models at +365 and +410 days indicate that there is likely a 
significant contribution from  [O~{\sc ii}] $\lambda\lambda$7320, 7330. If the blue peak  is indeed dominated by
[O~{\sc ii}] $\lambda\lambda$7320, 7330 then it is blue shifted by $-1920$~\kms\/ and has a width of around 1700~\kms\/ (from a 
blended two gaussian fit as evaluated by the 200d epoch). It is possible that this is [O~{\sc ii}]  from the same physical region that is producing the narrow  [O~{\sc iii}] lines, since the blueshift is similar. The velocity width of the line is lower than that estimated for 
[O~{\sc iii}] (of order 2900~\kms), but the blending of both these lines with the broad [Ca~{\sc ii}] component means that both estimates are uncertain. It is certainly clear that both are
narrower than the widths of main components of [Ca~{\sc ii}] and [O~{\sc i}], which are of order 7000-8000~\kms\/ and that the feature exhibits a blue-shift.

\begin{figure}
\includegraphics[width=\columnwidth]{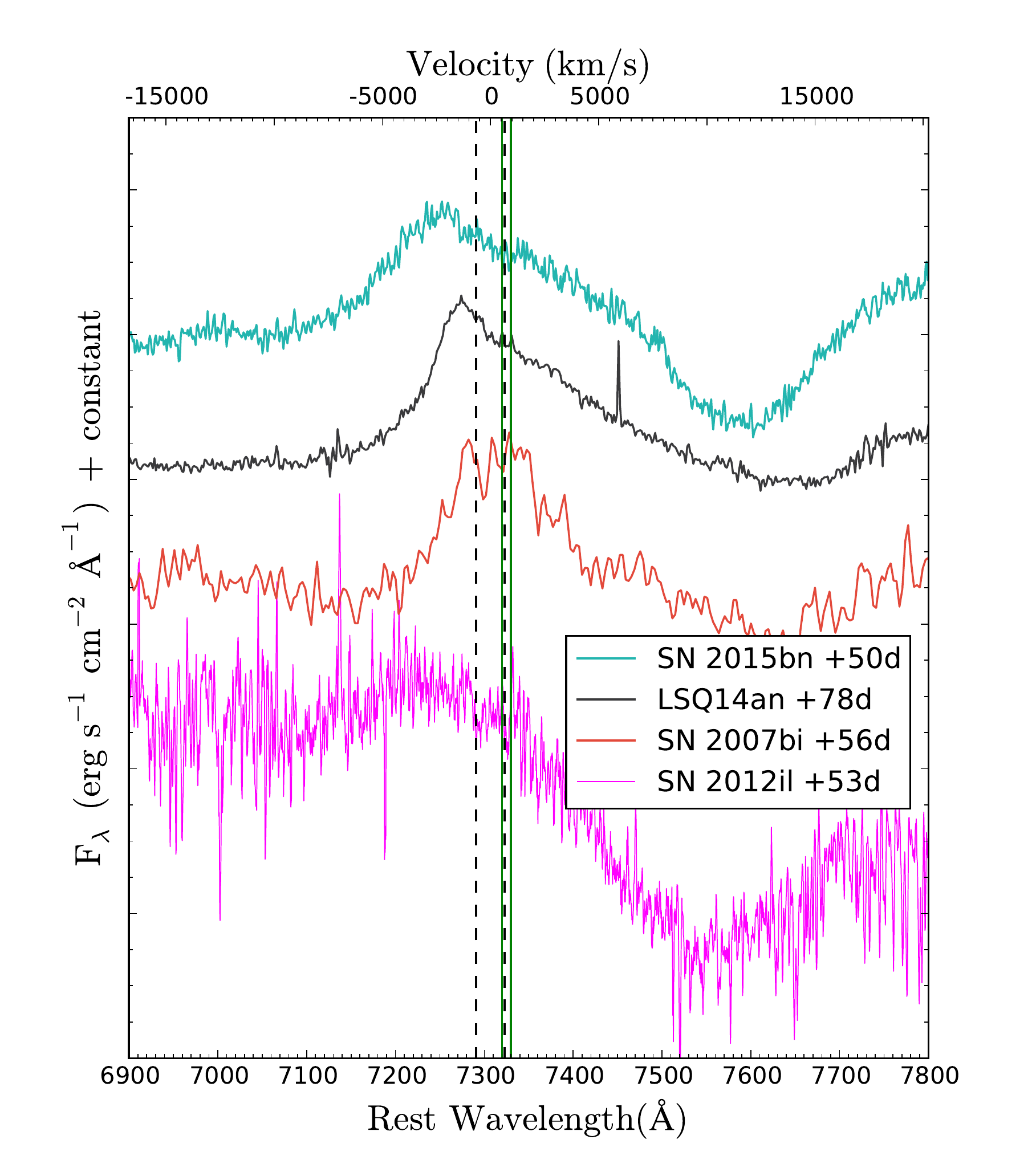}
\caption{Blue-shifted [Ca~{\sc ii}]+[O~{\sc ii}] profile of \an\/ compared to the same line of other two slow-evolving SLSNe at similar epoch. The two vertical black dashed lines denote $\lambda\lambda$7291, 7323 at rest-frame, while the solid green mark [O~{\sc ii}] $\lambda\lambda$ 7320, 7330.}
\label{fig:cac}
\end{figure}

Surprisingly, \an\/ is not the only slow-evolving SLSNe I showing this profile. As highlighted in Figure~\ref{fig:cac}, 
SN~2015bn\footnote{We chose an earlier spectrum of SN~2015bn, with respect to that of \an\/, due to its high S/N. However, the skewed [Ca~{\sc ii}] is also observable later than 50 days \citep[see][]{ni16a}.} 
shows a similar blue-shifted peak to that of \an\/, whereas that of SN~2007bi is centred at the rest wavelength. In PTF12dam, despite the low S/N of the spectra at that wavelength region, [Ca~{\sc ii}] also appears to be centred at the rest wavelength \citep[see extended data fig. 3 in][]{ni13}. Therefore we can conclude that the blue-shifted feature associated with [Ca~{\sc ii}]  $\lambda\lambda$7291, 7323 is not a unique feature of \an\/ but is also not ubiquitous in all slow-evolving SLSNe I.

A blue-shifted emission from [Ca~{\sc ii}] is not uncommon in core collapse SNe interacting with a CSM at late times.
Type II SNe 1998S, 2007od \citep{po04,in11} showed a similar blue-shifted [Ca~{\sc ii}] profile, as well as type Ib/c \citep[e.g.][]{ta09,mi10}. This is usually interpreted as a consequence of dust formation in ejecta, which causes the dimming of the red wings of line profiles due to the attenuation of the emission originating in the receding layers, or residual line opacity especially for stripped envelope SNe as also shown by models \citep[see][]{je15}. However, in the case of LSQ14an, 
only [Ca~{\sc ii}]+[O~{\sc ii}] and \oiii\/ show a blue-shifted peak and only [Ca~{\sc ii}]+[O~{\sc ii}] shows a clearly skewed profile.
If the explanation were dust, then it would suggest that the dust distribution is not homogenous and most likely in clumps. Hence 
we favour the component being either narrow [O~{\sc ii}]  arising from the same region as the [O~{\sc iii}]  lines (see Table~\ref{table:lines}), 
another component of [Ca~{\sc ii}] also located physically with the narrow [O~{\sc iii}] emitting region, or an effect of the residual line opacity. 
See Section~\ref{sec:dis} for more discussion and interpretation.

\section{Light and bolometric curves}\label{sec:lc}

\subsection{Light curves}
\begin{figure}
\includegraphics[width=\columnwidth]{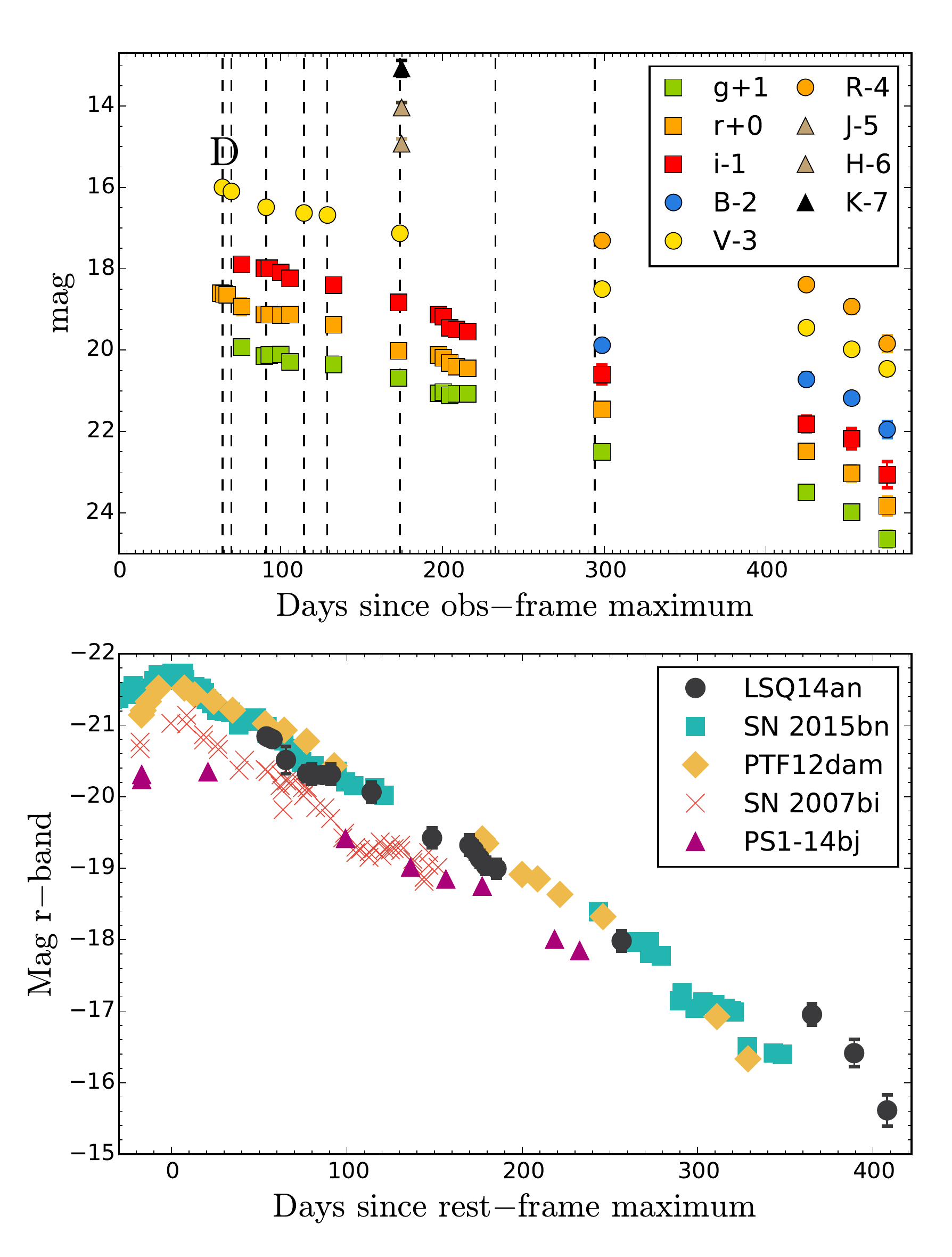}
\caption{{\it Top}: $griBVRJHK$ light curves of \an\/ in the observer frame with respect to the assumed maximum. The epochs of spectroscopy are marked with black, dashed lines, while the discovery with the letter 'D'. {\it Bottom}: Comparison of the $r$-band magnitude of well sampled slow-evolving SLSNe.}
\label{fig:lc}
\end{figure}

The \an\/ light curve shows a decline in all the available bands since the discovery epoch. This confirms that the SN was discovered past peak. However, the decline rate is not uniform and the uncertainty in the photometric measurements is small enough to 
detect clear changes in the decline. 
In the $r-$band we observe a 1.8 mag 100d$^{-1}$ decline (2.1 mag 100d$^{-1}$ in rest-frame) until 91 days past the rest-frame assumed peak, followed by a slower decrease of 1.2 mag 100d$^{-1}$ (1.4 mag 100d$^{-1}$ in rest-frame) from 114 to 257 days and a faster decrease of 2.7 mag 100d$^{-1}$ (3.1 mag 100d$^{-1}$ in rest-frame) in the last phase from 365 to 408 days. All declines are more rapid than that of \co\/ to \fe\/, which should be at $\sim$1.1 mag 100d$^{-1}$ in case of full trapping \citep{wb00}. 
In the bottom panel of Figure~\ref{fig:lc}, we show \an\/ $r$-band absolute light curve compared with those of four slow-evolving SLSNe with data during the same phase. All five are similar, with the resemblance between \an\/, SN~2015bn and PTF12dam 
being particularly close. This suggests that our assumed peak epoch from the spectra comparison should be correct within the uncertainties above stated. Previously SN~2007bi has been assumed, or proposed to be a unique object which requires
a physical interpretation as a pair-instability SN \citep{gy09}, but it is clear that there is a class of these slow-evolving SLSNe \citep[as initially proposed by][]{gy12},
which LSQ14an (and the others) are part. 

\begin{figure}
\includegraphics[width=\columnwidth]{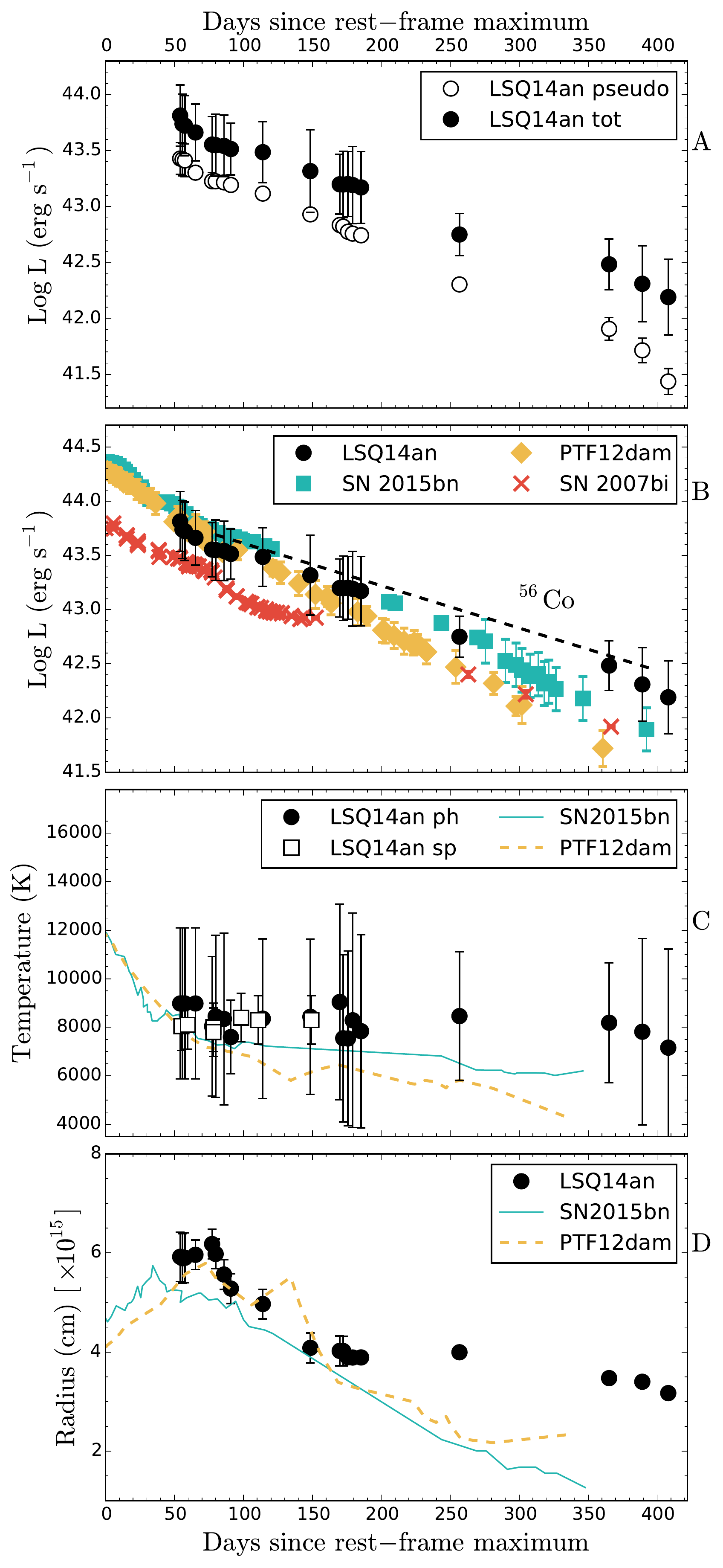}
\caption{{\it Panel A:} bolometric light curve of \an\/ obtaining by integrating the flux over the optical band (open circles) and from UV to NIR (black filled circles). {\it Panel B:} Bolometric light curve of \an\/ compared with those of SN~2007bi, PTF12dam, SN~2015bn. The \co\/ to \fe\/ decay slope is shown as a dashed black line. See text for further information. {\it Panel C:} temperature evolution of best-fit blackbody models evaluated for photometry (black circles) and spectroscopy (open squares) compared to that of the well observed SN~2015bn (cyan solid line) and PTF12dam (orange dashed line). {\it Panel D:} evolution of blackbody radius (black circles), compared to that of the same SLSNe of panel C.}
\label{fig:bol}
\end{figure}

\subsection{Luminosity, temperature and radius}\label{sec:bol}

Our \an\/ imaging dataset lacks UV coverage and has only one epoch of NIR coverage. Hence, only an optical pseudo-bolometric can be built \citep[see][for further details in the construction of the bolometric light curve]{in16b}. Applying the $K$-correction\footnote{In this paper all the absolute magnitudes have been $K$-corrected with the {\sc snake} code \citep{in16b}.}, fitting the available spectral energy density (SED) and integrating the flux from 1500 \AA\/ to 25000 \AA\/ should provide a good approximation to the full bolometric \citep[see][]{in13,in16a}. 
Pseudo and total bolometric light curves are shown in the top panel of Figure~\ref{fig:bol}, 
comparing the bolometric light curve of \an\/  with those of other slow-evolving SLSNe I, namely SN~2007bi, PTF12dam, SN~2015bn. They are all very similar, with decline rates from peak to 150 days that match 
the radioactive decay of \co\/ to \fe\/.  After 150 days, all of them deviate from the \co\/ rate, as already noted in the case of SN~2015bn by \citep{ni16b}. \an\/ is the less noticeable in this behaviour, possibly due to the sparse sampling in that phase combined with the light curve oscillations (see Section~\ref{sec:und}). If 
they were mostly powered by \ni\/  then they can not be fully trapped. This argues strongly against them
being pair-instability explosions, since the massive ejecta required for this explosion mechanism inherently 
implies full trapping up to 500 days after maximum light \citep{je16a}. Hence, while at first sight the 
approximate match to \co\/ is appealing to invoke nickel-powered explosions, it quantitatively does not fit with 
massive pair-instability explosions. We also see some evidence that after  250--300 days, there may be  a steepening in the decline slope in all four. The decline at times after 300 days
is roughly consistent with a power-law of index $\alpha = 5$ ($L\sim t^{-5}$), typical of an adiabatic expansion (a.k.a. Sedov-Taylor phase), which is unexpected at this phase (see Section~\ref{sec:dis}).

The blackbody fit to the photometry delivers information about the temperature and radius evolution. We also used a blackbody fit on our spectra in order to increase the available measurements. We then compared our results with the well sampled PTF12dam \citep{ch15,vr16} and SN~2015bn \citep{ni16a,ni16b}. \an\/ shows an overall steady temperature of 8000K, which is slightly higher than that measured in SN~2015bn and PTF12dam but still consistent within the errors. 

The evolution of the blackbody radius of  \an\/ shows an overall similar behaviour to that of PTF12dam and to SN~2015bn up to 150 days. After 150d it remains roughly constant until the end of our dataset in a similar fashion to that of PTF12dam, while SN~2015bn steadily decreases. A small increase of the radius is noticeable from $\sim$60d to $\sim75$d, 
which corresponds to the first of the two possible fluctuations observed in \an\/ light curve. Such behaviour was also observed for the undulations in SN~2015bn \citep{ni16a}.

\section{Light curve undulations in hydrogen-poor slow-evolving SLSNe~I}\label{sec:und}

\begin{figure}
\includegraphics[width=\columnwidth]{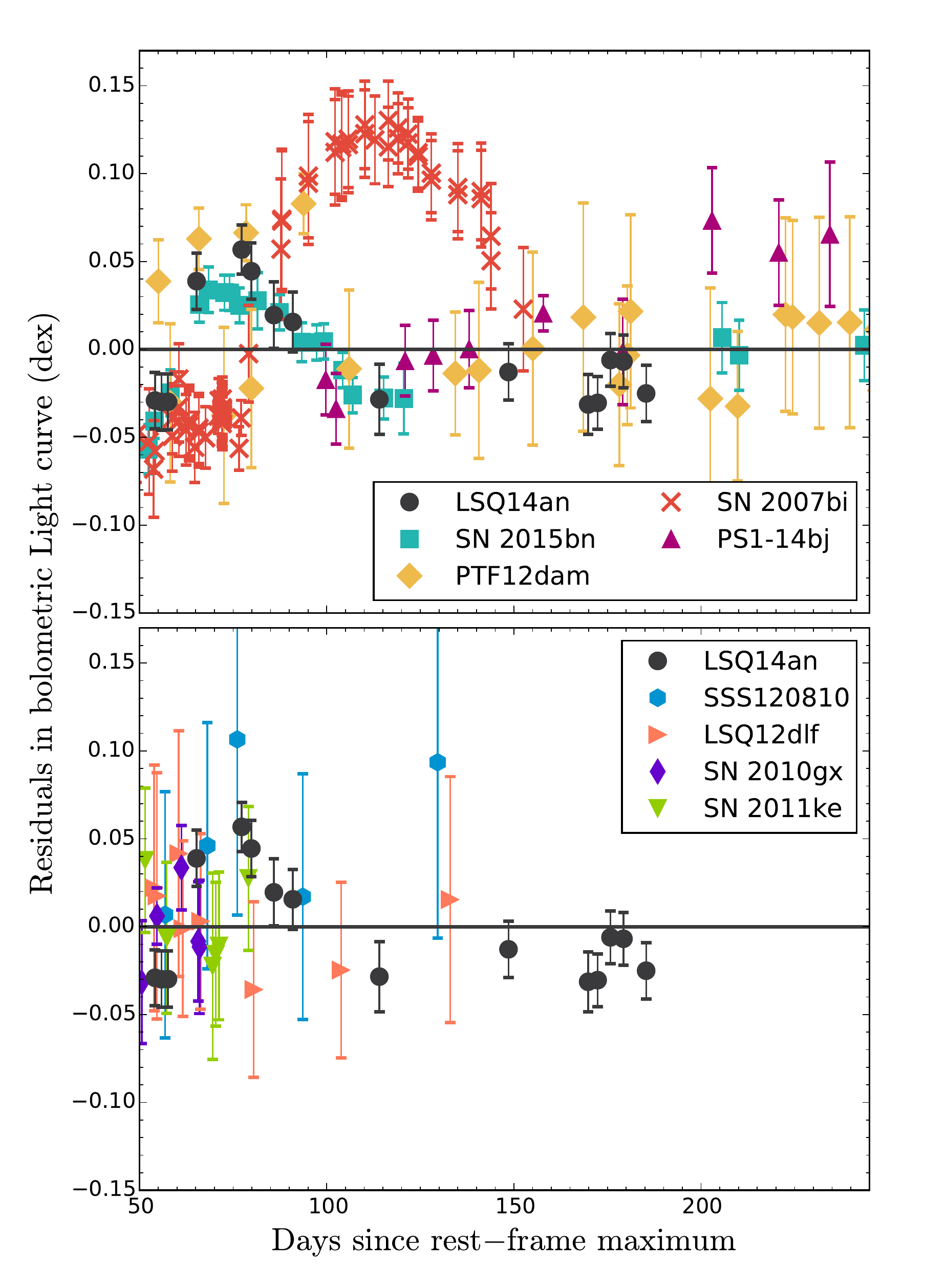}
\caption{{\it Top:} Residuals, after subtracting fits to the declining bolometric light curve, showing undulations over the selected phase in the case of slow-evolving SLSNe. {\it Bottom:} LSQ14an residuals compared with those of four fast-evolving SLSNe, which do not show any appreciable fluctuations, with the possible exception of SSS120810.}
\label{fig:res}
\end{figure}

All the individual \an\/  light curves show some low level evidence of undulations. To explore this further, we fitted a first order polynomial to the bolometric light curve. The choice of using the bolometric allows us  to avoid an analysis dominated by $K$-correction or line evolution within a particular pass-band. We chose to focus  from 50 to almost 250 days excluding data after 250 days due to sparse sampling. 
 The choice of a linear function to fit is due to the fact that slow-evolving SLSNe I light curves are usually well reproduced by models having a uniform, steady decline over the analysed baseline. We then subtracted the fit from the light curve over that timescale. The residuals are shown on the bottom panel of Figure~\ref{fig:res}. 

The residuals appear to have at least one clear fluctuation pattern from 50 to 200 days in rest-frame. At 75 days we observe a fluctuation of 0.06 dex from the linear fit.  Another less distinct oscillation (0.04 dex) is visible around 170 days. Nevertheless, there is a residual pattern which oscillates across the linear decline fit with an amplitude that appears larger than the errors on the flux measurements. 

We then compared the \an\/ residual analysis with those of our selected slow-evolving SLSNe I sample. The phase analysed here is after that of the `knee' observed in SN~2015bn \citep{ni16a}. We also observe a small fluctuation from 75 until 120 days in  the SN~2015bn data, similar to that of \an\/ but with lower amplitude (0.02 - 0.03 dex). Nonetheless,  our estimated errors are smaller than the observed amplitude for this SN too.  PTF12dam may also have some undulations present,
 but in this case the amplitude is comparable or less than the errors. Fluctuations are also observed for PS1-14bj, which shows a 0.03 dex fluctuation at around 100 days and a more noticeable one (0.07 dex) around 220 days. For all  objects the undulations are small, with an average of 0.05 dex, but they are  generally greater than the photometric errors (0.01-0.03 dex depending on the phase and distance of the object). On the other hand, SN~2007bi shows stronger oscillations from 0.05 (at 60 days) to 0.12 dex (at 115 days), a factor ten bigger than the errors. Following on from the analysis of SN~2015bn by 
\cite{ni16a} which first highlighted these undulations, it may be that 
light curve fluctuations are common place in slow-evolving SLSNe I, but that their magnitude is highly variable from 
object to object. 
On the other hand, we also checked if such behaviour was present in the fast-evolving SLSNe I. We used data of the well sampled SNe 2010gx and 2011ke \citep{pa10,in13} together with LSQ12dlf and SSS120810, which have data up to $\sim$140d \citep{ni14}. We used a linear fit in the case of SN~2010gx, LSQ12dlf and SS120810 and a second order polynomial in the case of SN~2011ke. The difference in the latter is due to the fact that the SN light curve undergoes a noticeable transition to 
 the tail phase just around $\sim$50 days \citep[see][]{in13} and a linear fit would have produced a bogus oscillation in the residuals. SSS120810 seems to experience a genuine oscillation around $\sim$70 days, as already noticed by \citet{ni14}, but the rise is sharp and the width is small, especially when compared with the slow-evolving. LSQ12dlf could have experienced an oscillation similar in amplitude to those of before, or alternatively could be the beginning of the tail phase, but the errors and the lack of later data prevent any conclusion.
In general, we do not see any clear trend in the residuals resembling those of slow-evolving SLSNe I, even though the time coverage is too short and future work should analyse earlier epochs (e.g. $0<$ phase (d) $<60$) where fast-evolving SLSNe are better sampled. 

In the magnetar scenario, light curve oscillations could be the consequence of the magnetar ionisation front. In particular, when O~{\sc ii} and O~{\sc iv} layers driven by the hard radiation field of the magnetar reach the ejecta surface a change of the continuum opacity will happen \citep{me14}. Lower ionisation states generally penetrate further and reach the surface earlier. The less ionised layer should reach the surface in $\sim10$ days from explosion (well before our detections of the 
light curve deviations). The other should happen at the same time of an X-ray shock breakout that, considering the ejecta masses and spin period of slow-evolving SLSNe I, should happen more than a year after peak (see Section~\ref{sec:xr}).
Hence, a sudden rise in opacity,
would be non trivial to account for and would require a radiative transfer calculation to test the feasibility of this scenario. 
Multiple changes in opacity, which would be required for SNe 2007bi and 2015bn, would be more difficult to explain. 

Another option suggested to explain the pre-peak undulation of SN~2015bn \citep{ni16a} is that the fluctuation is powered by oxygen recombination, which would be below 12000~K \citep{hat99,qu13,in13}, in a similar fashion to what occurs in type II SNe within the hydrogen layer. Since the oscillations after 50 days already happen at an almost constant temperature of T$_{\rm off}\approx8000$K (see Section~\ref{sec:bol}) it would be difficult to explain all of them with this interpretation.

On the other hand, oscillations have been observed in SN light curves of  interacting transients such as
type Ibn, IIn or SN impostors \citep[e.g. SNe 1998S, 2005ip, 2005la, 2009ip, 2011hw; SNhunt248][]{fa00,sm09,fox09,st12,pa08,fr13,mar14,ma15,sm12,pa15,erkki15}. To reproduce the short time scale oscillations observed in slow-evolving SLSNe (amplitude of 50-100 days) with ejected masses of 7-16 \M\/ \citep{ch15,ni16a,in16c} may be difficult since massive ejecta should not produce small scale changes in the light curve \citep[e.g.][]{gaw10,fr13}. This, together with the small increase in the radius, would disfavour an interaction scenario such as that of pulsational pair instability SNe (PPISNe) where thermonuclear outbursts due to a recurring pair-instability release massive shells of material will collide with each other if the latter shells have velocity higher than the initial \citep{woo07}. As a consequence, a viable scenario would need a relatively low mass CSM ($<$1 \M) as that proposed  by \citet{mo15} for the late interacting slow-evolving SLSN iPTF13ehe \citep{yan15}. Indeed, assuming an average half-period of the fluctuation t=25 days, average luminosity L=10$^{43}$ erg s$^{-1}$ and $v$=7000 \kms\/ \citep[this work,][]{gy09,yo10,ni13,ni16a} and using the scaling relation ${\rm L} \approx {\rm M_{CSM}}v^2 / 2{\rm t}$ \citep{qu07,smc07} we find that a CSM of no more than M$_{\rm CSM}\sim0.04$ \M\/ could be responsible for each fluctuation, similar to the findings of SN~2015bn \citep{ni16a}. If the light curve oscillations are periodic, they could be the consequence of a close binary system that drives the stripping of the progenitor stars (or the companion) causing a heterogeneous density structure of the CSM \citep[e.g.][]{we92,mo15}.

\begin{figure*} 
\includegraphics[width=18cm]{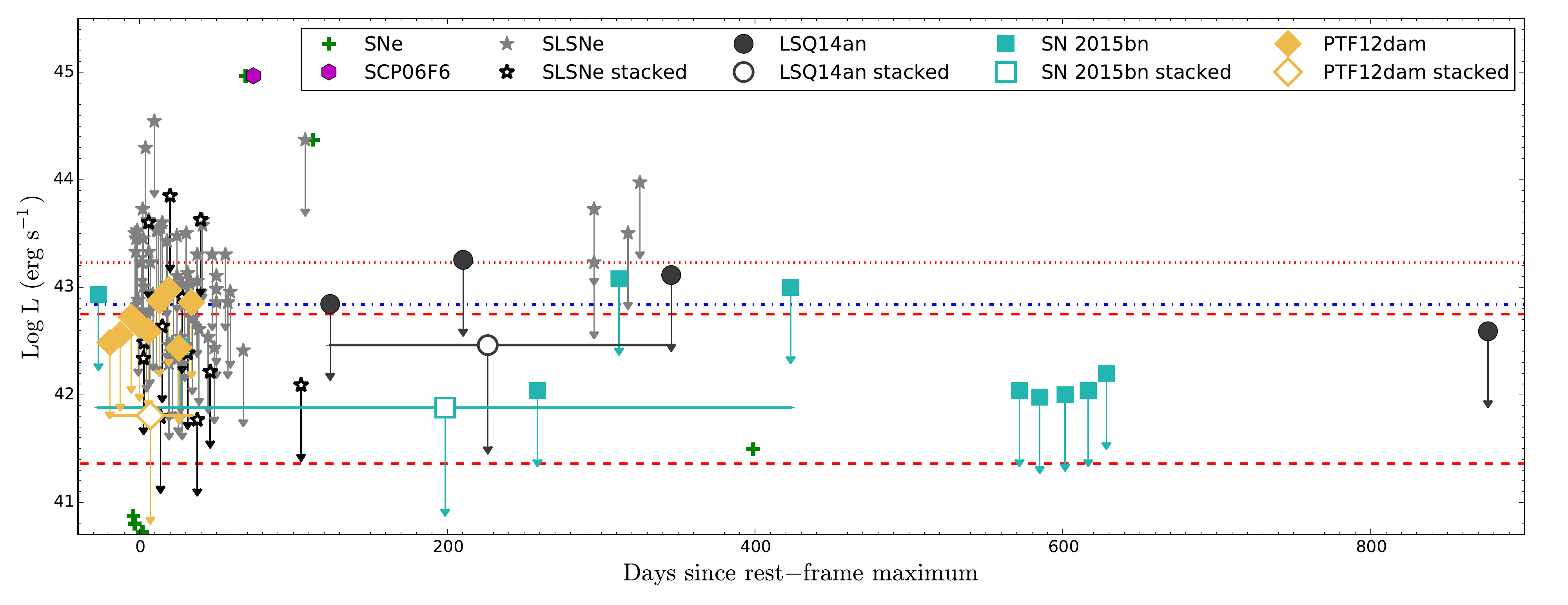}
\caption{SLSNe I X-ray luminosity limits (grey stars) and stacked limits (black open stars) as reported by \citet[][but evaluated using the cosmology adopted here]{lev13}, compared with those of the available slow-evolving SLSNe~I \an\/, SN~2015bn (this work) and PTF12dam \citep[][]{lev13}, the single X-ray detection of SLSN I SCP06F6 and a sample of SNe detections \citep{of13}. X-ray limits in case of break-out from the pulsar nebula ionisation front (blue dashed dotted line) and interaction (red dashed lines for the two non massive CSM cases and red dotted for the massive CSM scenario) are also shown. See text for further details.}
\label{fig:x}
\end{figure*}

\section{X-ray limits in slow-evolving SLSNe~I}\label{sec:xr}

An alternative method to investigate the SLSNe progenitor scenario and powering mechanism is that of X-ray observations. Both the magnetar and the interaction scenario would predict X-ray emission but at different luminosities and at different phases.

In the case of the magnetar scenario, the pulsar wind inflates a hot cavity behind the expanding stellar ejecta, which is the pulsar wind nebula. Electron/positron pairs cool through synchrotron emission and inverse Compton scattering, producing X-rays inside the nebula. These X-rays ionise the inner regions of the ejecta, driving an ionisation front that propagates outwards with time \citep{me14}.
\an\/ shows a light curve similar to that of SN~2015bn and hence we can assume similar magnetar and ejecta parameters to those retrieved by \citet[][P = $1.7-2.1$ ms, B$_{\rm 14}=0.9-1.0$ G and M$_{\rm ej}=8-15$\M\/]{ni16a} in their magnetar fit \citep[for the magnetar semi analytic code see prescription of][]{in13}\footnote{The code is available at \url{https://star.pst.qub.ac.uk/wiki/doku.php/users/ajerkstrand/start}}. The mass range is similar to that
 estimated by other work on slow-evolving SLSNe \citep{ni13,ch15}, with the only exception of PS1-14bj \citep{lu16} showing redder spectra, longer rise-time and fainter peak luminosity than the bulk of slow-evolving SLSNe I. Thus, for our analysis, we can use an average mass of 12~\M\/ for \an\/, SN~2015bn and in general for all slow-evolving SLSNe I.
Following the prescription of \citet[][their equations 5, 6, 56 and A11]{me14} we would expect an X-ray break-out of L$_{\rm X}$~$\sim7.0\times10^{42}$ erg s$^{-1}$ at t$_{\rm X}>700$\footnote{We note that these calculations are highly sensitive to the ejecta mass, that however seems similar for all slow-evolving SLSNe I.} days from explosion that corresponds to $\gtrsim$630 days from maximum light for an average rise-time of about 70 days.

Alternatively, in the interaction scenario the shock waves created by the interaction between the SN ejecta and the CSM heat gas to X-ray emitting temperatures and can possibly accelerate particles to relativistic energies. The presence of undulations in the light curves, would disfavour the presence of a massive circumstellar medium \citep[7-19 \M\/;][]{ch13,ni16a}, which would also power the peak luminosity \citep{ci11}. In contrast, a less massive circumstellar medium ($<$1 \M, see Section~\ref{sec:und}) could be more suited to explain \an\/ and other SLSNe I observables.
Generally, during the interaction process between SN ejecta and CSM, the conversion of the kinetic energy into radiation is affected by complicated hydrodynamic and thermal processes including thin shell instabilities \citep{vi94}, the Rayleigh--Taylor
instability of the decelerating cool dense shell, the CSM clumpiness, mixing, and energy exchange between cold
and hot components via radiation and thermal conductivity \citep{cha15}. Here we use a simplified approach in which the X-ray luminosity of both shocks is equal to the total kinetic luminosity times the radiation efficiency \citep[$\alpha_{\rm X}>0.1$;][]{mo13,cha15}. As a consequence, the energy mass relation would be E~$\propto {\rm M}^{0.7}$, assuming an outer power law of the density of the ejecta n=10.

This is equal to L$_{X}\sim2.3-57.3\times10^{41}$ erg s$^{-1}$ for a 0.04-1.0 \M\/ CSM interacting for 200 days (the time when we observe the fluctuations) and to L$_{X}\sim1.7\times10^{43}$ erg s$^{-1}$ for 13 \M\/ interacting for 400 days, which corresponds to the massive CSM case. In both cases we assumed $\alpha=0.1$ and an ejecta always more massive than CSM \citep[e.g.][]{ch13,ni14}. We note that in interacting SNe there is a higher efficiency in the optical \citep[where $\alpha\geq0.5$;][]{mo13} than in X-ray \citep{of14,cha15}, hence the highest luminosity achievable in our case would be five times those above. On the other hand, to estimate our X-ray luminosities we did not consider any photoelectric absorption. In our case such contribution is hard to evaluate and varies with time, decreasing the soft X-ray output after the start of interaction \citep[see][]{ch03}. This absorption would increase in case of massive CSM with respect to normal interacting SNe (CSM$\lesssim$1\M), but the X-ray emission would increase too. Hence, the values reported above have to be treated as higher limits of L$_{X}$ and they should be similar to the X-ray luminosity emitted around maximum optical light, or soon before that, in case of interaction with a massive CSM \citep[see][]{ch03}.

We report here the measurements for \an\/ and SN~2015bn spanning 124--876 days from assumed maximum for the former, and from -27 to 628 days for the latter. In order to place meaningful constraints, we chose observations having at least 4000s exposure or covering the late phase of the light curve ($>$300 days since maximum). To date, no observation has been carried out later than those reported  here and no X-ray emission has been detected in the data here reported, neither for the stacked images of each SN (evaluated adding all the frames together and assuming an average phase, see Table~\ref{table:xr})\footnote{For both SNe we only stacked the first three epochs since it would have been deceptive to also add the later one, which were carried out from five months to more than a year after the end of our optical campaign.}. For both SNe we assumed the H-column density to the Galactic value (n$_{\rm H}$(\an)~=~4.16~$\times$~10$^{20}$~cm$^{-2}$ and n$_{\rm H}$(SN 2015bn)~=~1.22~$\times$~10$^{20}$~cm$^{-2}$) since SLSNe are hosted in galaxies with little to no internal reddening \citep[see][]{lu14,le15,pe16}.  We used a generic spectral model with a photon index $\Gamma=2$ \citep{lev13} to estimate the luminosity limits. We note that limits can change by a factor of $\sim$3 in flux for a variation of the photon index of order of ten. We also used a thermal model with $kT=1.55$  \citep[equal to that used for SCP06F6;][which is the only SLSN detected in X-ray]{lev13} and hence a factor of 1.2 to 10 lower than those used in H-rich interacting SNe \citep[e.g.][]{cha12,of14,cha15}.

Figure~\ref{fig:x} shows the average limits between the two methods with the power law method usually resulting in higher luminosities than the thermal one, but still on the same order of magnitude. Only the stacked limits, \an\/ limit at 876d, SN~2015bn limit at 259d and limits at $>$571d, are below the magnetar threshold reported above (dotted dashed blue line in Figure~\ref{fig:x}). SN~2015bn phase is from $\sim400$ to $\sim$30 days earlier than that of the expected ionisation front break-out, while \an\/ is $\sim300$ days after and the break-out does not last so long at that luminosity. Indeed it declines with time \citep[see][]{me14}. 
However, one needs to be very careful in interpreting the stacked limits in a meaningful way. They are only 
quantitatively meaningful if they are taken to represent a constant level of flux across the epochs which are stacked. In 
other words they by no means rule out time variable flux between the observed points. 
All limits are below the X-ray luminosity (dotted red line) that we should have observed in case of interaction with a dense and massive CSM or collision between two massive shells as suggested by the PPISNe scenario. However, in the case of a massive CSM, the X-ray luminosity should decrease from peak epoch due to photoelectric absorption \citep{ch03} and after a year should be a factor of two to fifteen less than that at peak. Hence, the late time limits could not be so constraining\footnote{The photoelectric absorption strongly depends on the column density of the cool gas behind the reverse shock. This is difficult to evaluate in a case of a massive ($\gtrsim$10\M) H- and He-free (or deficient) shell in a sub-solar metallicity environment since it would require assumptions on each term of the formulas. Hence, we avoid any proper calculation of this effect but we would like to warn the reader that photoelectric absorption cannot be ignored and would somehow decrease the X-ray emission at late time.}. On the other hand, in case of interaction with a less massive CSM (0.04 \M, lower dashed red line) we would have expected an X-ray emission at a similar phase to the fluctuations and hence earlier than 250 days. Such X-ray would have luminosities lower than our limits. This is fulfilled for both limits and stacked limits and all slow-evolving SLSNe. Although we have made many simplifying assumptions, 
the most simple interpretations is that the 
X-ray limits  disfavour the scenario in which the interaction of ejecta 
with a massive CSM is the powering mechanism for this SLSN. 
However both the magnetar engine and interaction with a less massive CSM could play a role in the spectrophotometric behaviour of slow-evolving SLSNe~I.

\section{Discussion on the characteristics of slow-evolving SLSNe I}\label{sec:dis}

The \an\/ dataset is limited to post-peak, but the density of the spectra and light curve monitoring and the two excellent X-Shooter 
spectra offer interesting further information to better understand the explosion scenarios of slow-evolving SLSNe I.
In summary it is a slow-evolving SLSN of type I or Ic. There are a group of, at least, four low redshift (z$<$0.2) SLSNe I (or SLSNe~Ic) that show very 
similar spectrophotometric evolution SN~2007bi, PTF12dam, SN~2015bn and LSQ14an. They have very broad
light curves that initially ($<150$d after peak) decline at a rate similar to $^{56}${Co}, although it is unlikely that this is the true power source of the luminosity.  They have higher redshift analogues such as LSQ14bdq \citep[$z=0.3450$;][]{ni15a}
PS1-11ap \citep[$z=0.5240$;][]{mc14}, DES13S2cmm \citep[$z=0.6630$;][]{pap15}, iPTF13ajg \citep[$z=0.7403$;][]{vr14}, SN 2213-1745 and SN 1000+0216 \citep[$z=2.0458$ and $z=3.8993$, respectively;][]{2012Natur.491..228C}.  PS1-14bj \citep[$z=0.5215$;][]{lu14} is also similar but has the most extreme in terms of its light curve width, while iPTF13ehe \citep[$z=0.3434$;][]{yan15} shows a late time re-brightening due to interaction with a H-shell (something similar could have happened also to LSQ14bdq, Schulze private communications). 
In Sections~\ref{sec:o3}~\&~\ref{sec:ca2} we reported two unusual characteristics of the spectra of LSQ14an, some of which had been observed before, as well as a somewhat common feature in the spectral evolution of this group of SLSNe. Furthermore, in Sections~\ref{sec:lc}~\&~\ref{sec:und} we described distinctive features of \an\/ light curves that seem shared by the sample of slow-evolving SLSNe~I.

\subsection{Spectroscopic characteristics}

The first spectroscopic characteristic is the presence of \oiii\/ lines which are
significantly broader than the host galaxy, unambiguously formed in the SN ejecta and blueshifted compared to the rest of the ejecta (see Section~\ref{sec:o3} and Table~\ref{table:lines}). 
The second is the existence of a strong blueshifted component 
associated with the prominent [Ca~{\sc ii}] line (see Section~\ref{sec:ca2}). 

In \an\/ and PS1-14bj, the \oiii\/ lines have $v\sim3000\pm300$ \kms\/. 
In the case of \an, this is significantly lower than that of the bulk of the ejecta as traced by the strong [O~{\sc i}] and O~{\sc i} lines, around 8000~\kms 
(Mg~{\sc i}] is broader at $\sim$12000~\kms, although it may be a blend).
Furthermore, the density of the \oiii\/ emitting region is likely much less than that where [O~{\sc i}] is formed. This would suggest multiple emitting regions for the above ions. Ionised elements show line profiles  much narrower than the neutral ones, strengthening the hypothesis that they come from different regions and likely from a region interior to that where the neutral lines are formed. Since [O~{\sc i}] travel faster than \oiii\/ (see Section~\ref{sec:o3} and Table~\ref{table:lines}), another possibility is to have a fast-dense outer region and a slow-diffuse region similar to an evacuated inner cavity, which could be another indirect evidence of a magnetar inner engine. 

In the magnetar or inner engine scenario, these  blue-shifted lines and multiple emitting regions could be explained if the ejecta was fragmented or axis-symmetric and the observer was in the direction of the axis of symmetry (or both). The latter would happen since only the nearer (blue) component is observed, while red photons would be scattered perpendicular to the observer.
An axisymmetric ejecta has been observed for SN~2015bn \citep{in16c}, while a certain degree of fragmentation, and also asphericity of the ejecta has been observed in 2D magnetar models \citep{chen16}.
On the other hand, if we assume a certain degree of interaction with a small amount of material (see Section~\ref{sec:und}), the blue-shift could be due to an additional contribution to the line coming from the cool dense shell (CDS), consequence of the shock propagating in the ejecta. This is blue-shifted because the CDS is moving at a velocity of $\sim1500$ \kms\/ and is optically thick in the continuum, thus line photons emitted in the red are absorbed by the near side of the CDS \citep{de15}. Since a high dilution factor is required to model [O~{\sc iii}] and [O~{\sc ii}] \citep{je16}, this could be achieved in a diluted interaction region.
We already mentioned as [O~{\sc ii}] and [O~{\sc iii}] lines are much narrower than neutral lines, which in the simplest scenario means an origin in the inner, slower ejecta. Lines from the inner ejecta are normally more prone to obscuration by a residual opacity (a residual ``photosphere"), which can also explain their suppressed red wings. This would be another piece of circumstantial evidence for a hot, ionised inner region and thereby central powering.

Finally, an important point is that the spectra of slow-evolving SLSNe~I at 50--70 days, surprisingly show 
almost identical emission features to those seen in the nebular phase at 300-400 days. The spectra at each phase
have quite different continuum properties (the early spectra being much bluer), but the ionic transitions are the same. 
 [O~{\sc i}] $\lambda\lambda$6300, 6363 significantly increases in strength with time (as expected in CC-SNe), but
others such as [Ca\,{\sc ii}] and [O\,{\sc iii}] are relatively constant in luminosity. 
This implies that the line forming regions of these lines have values of density, temperature and opacity that 
remain roughly constant over this extended period of evolution. That could also be a consequence of a nearly constant powering mechanism.
We see that the emission lines which are redder than 5500~\AA\/ show a shift in the centroid of the peak 
between their values at 
50--70 days  and those at $>$300 days (see Figure~\ref{fig:cmpel}). For example in SN~2015bn the [O~{\sc i}] and O~{\sc i} centroids move toward bluer wavelengths, while [Ca~{\sc ii}] moves to the red. 
Multiple emitting regions may thus be present not only in \an\/ but may be a more common feature in slow-evolving SLSNe I. If slow-evolving SLSNe~I are the more massive version of the fast-evolving as suggested by \citet{ni15}, sharing similar explosion energies and the same mechanism responsible for the enormous luminosities, we would expect to observe nebular lines later than the fast-evolving, as observed for normal stripped envelope. For example  in SN~2009jf, the onset of 
emission from forbidden calcium is delayed until  85 days from peak \citep{va11}. 
However, nebular emission lines (e.g. [O~{\sc i}] or [Ca~{\sc ii}]) in fast-evolving SLSNe I have only been observed at $\sim150$ days after peak \citep{kan16}\footnote{We note that \citet{kan16} reported that SLSN Gaia16apd shows a fast decline at early phase similar to SNe 2010gx and 2011ke but a late ($>$130d) behaviour between the fast- and slow-evolving events.} and not before $\sim100$ days \citep{qu11,in13}. 

Another important characteristic shown by slow-evolving SLSNe~I is that  their early time spectra are always bluer than the fast-evolving and the late time spectra are still remarkably blue even past 200 days
 \citep[see fig.~7 in][and Figure~\ref{fig:cmpel} here]{je16}. They are significantly  bluer than stripped envelope SNe, or 
the faster evolving SLSNe as also noted by \citet{ni15,ni16b} for SN~2015bn. That could be explained if we assume a
fairly constant $8000\pm1000$~K blackbody type of emission with the observed forbidden and semi-forbidden lines 
in addition (as seen in the photometric fits in Figure\,\ref{fig:bol}).  This type of underlying continuum emission is 
common scenario in  SNe dominated by interaction, however  in slow-evolving SLSNe~I we still lack the presence of narrow lines forming in the un-shocked material. Alternatively, this could be explained by multiple optically thick lines creating a blackbody-like continuum with the forbidden and semi-forbidden lines forming in more external region than those creating the pseudo blackbody continuum. As highlighted by \citet{je16}, a dense CSM shell of $\sim10$\M\/ could reproduce the multiple emission regions and the low filling factor required, but published models are still not able to reproduce the bulk of velocities and the absence of narrow lines.

\begin{figure} 
\includegraphics[width=\columnwidth]{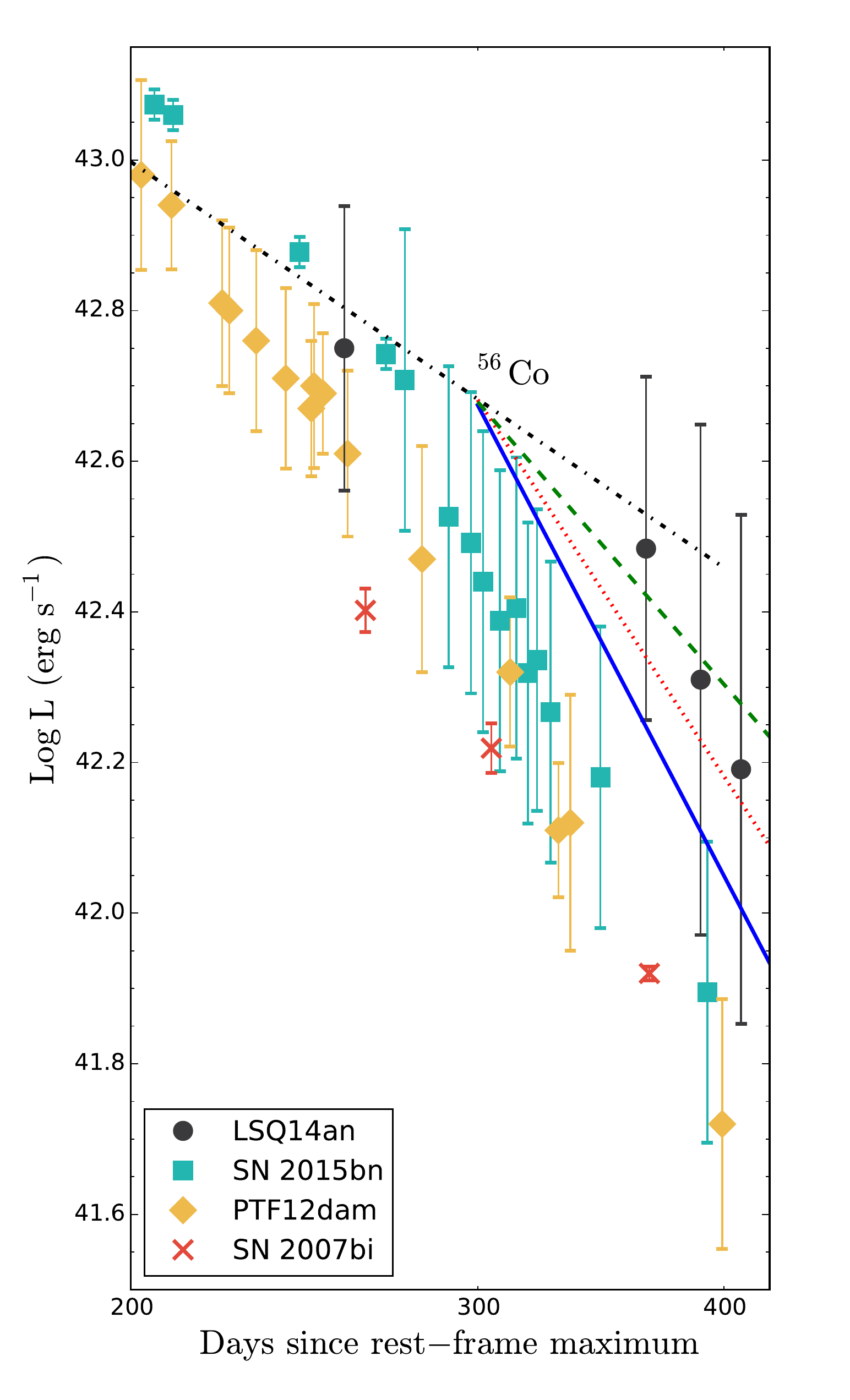}
\caption{The late time bolometric evolution of \an\/, together with those of slow-evolving SLSNe I SN~2007bi, PTF12dam, SN~2015bn. The green dashed (n = 3) and red dotted (n = 4) lines show the decline similar to a snowplow phase, while the solid blue line that similar to an adiabatic (Sedov--Taylor) phase. The black dot-dashed line shows the expected \co\/ decay. The x axis is in semi-log scale to highlight the decline and the differences in the power law indexes.}
\label{fig:late}
\end{figure}

\subsection{Light curve characteristics}

We also highlighted that while the light curve declines 
at a rate similar to $^{56}${Co} until 150 days, it increases (for all the objects) from 150 days onwards with a possible further steepening after $\sim300$d. The change in decline after 150 days disfavours the pair instability scenario for slow-evolving SLSNe~I since PISN explosion should exhibit full gamma-ray trapping and hence closely follow the decay of \co\/ to \fe\/ up to 500 days \citep{je16a}. Furthermore, there are some likely real undulations (see Section~\ref{sec:und}) in the bolometric luminosity. 
These peculiarities could be present in the higher redshift analogues, but a combination of lower quality data 
at red wavelengths (observer frame) and late time  (rest-frame) due to their high redshift prevent a proper investigation.

First of all we consider the possibility that these features are a consequence of the magnetar scenario.
The only possibility would be a change in the density of the ejecta due to the magnetar ionisation front, linked to oxygen ionised states \citep[mainly O~{\sc ii} and O~{\sc iv},][]{me14}, and moving toward the surface of the ejecta. This happens twice but the timescale is not consistent for both since the former occurs 3--4 weeks after the explosion, while the other at least 700 days after explosion. Moreover the second is also coincident with an X-ray breakout that has not been observed during the time frame in which we observed the undulations.
Secondly, we examine the possibility that the phenomenon is due to some ejecta - CSM interaction. However even 
if it is, it is unlikely that this CSM interaction powers the full luminosity of the light curve since our X-ray analysis (see  Section~\ref{sec:xr} for the assumed caveats) does not support the interaction with a massive CSM shell ($\sim$13 \M).

We also see some evidence of an increase in the decline rates after 300 days (see Figure~\ref{fig:late}). This is observed in \an\/, SN~2015bn and PTF12dam and possibly even SN~2007bi, even though for this object the increase is less certain. This decline is steeper than $t^{-3}$ or $t^{-4}$, which is known as the snow-plow phase and it has been observed in H-rich interacting SNe \citep[e.g.][]{of14,fr14,in16a}. 
Figure~\ref{fig:late} shows that the decline observed in slow-evolving SLSNe I after 300 days seems to follow a power law of $t^{-5}$, which is the same slope of an adiabatic expansion phase (the Sedov-Taylor phase). This would suggest conservation of energy that, at this time, could be achieved when the density is low enough and the gas cannot cool quickly. An adiabatic expansions at late time of a SN evolution happens when the mass of gas swept up becomes greater than the mass of ejecta and the kinetic energy of the original exploded envelope is transferred to the swept up gas, and that is heated up by the shock wave. This would require a large amount of interstellar medium or CSM. However, spectra and X-ray analysis of the sample do not show any of the {\it primary} observables usually seen in interacting SNe (e.g. narrow lines or X-ray emission).

In the case of \an\/,  considering a hypothetical rise time from explosion to peak  of approximatively 70 days \citep{ni13,ni16a} and using the simple formalism of R$_{\rm shock} =2.1\times10^{13}\, (E/\rho)^{1/5}\, t^{2/5}$ \citep[][]{re08}, where t = 370 days from explosion and $\rho\sim10^{8}$ \citep[][and Section~\ref{sec:o3}]{je16}, we derive a distance of 1.2$\times10^{23}$ cm for the radius of the shock, which is at least a factor of 10$^8$ higher than the SN radius. This suggests it is too early for SLSNe to experience a ``natural'' adiabatic phase, that indeed usually happens after 200yrs in normal supernovae \citep{re08}. 
 The approximate calculation above is also valid (in terms of rough values) for 
 SN~2015bn and PTF12dam. 
A possible alternative is that these slow-evolving SLSNe experience a further decrease of the $\gamma$-ray trapping \citep[see also][]{ch15,ni16b}. A further investigation is needed to understand if this behaviour belongs to all slow-evolving SLSNe~I and what is
the physical reason. We note that none of the fast-evolving SLSNe~I discovered up to date has observations at $>300$ days, which prevent a similar investigation.

\section{Summary and deductions}\label{sec:conc}

We have presented an extensive dataset on 
\an, covering its spectrophotometric behaviour from 55 to 408 days together with X-ray limits spanning over the same timescale and up to $\sim$870 days.  We show that there are now four well observed, low redshift SLSNe~I which are similar in their photometric
and spectroscopic evolution: SN2007bi, PTF12dam, SN2015bn and LSQ14an. The combination of the datasets provides
some new insights into these SNe.

LSQ14an was classified after maximum light and exhibits spectroscopic peculiarities such as broad ($v\sim3500$ \kms), blue-shifted \oiii\/ $\lambda\lambda$4959, 5007 lines, a blue-shifted peak of [O~{\sc ii}] $\lambda\lambda$7320, 7330 (or [O~{\sc ii}] + [Ca~{\sc ii}] $\lambda\lambda$7291, 7323), as well as the late appearance of NIR Ca and the remarkable fact that the lines observed at late time (300-400 days after maximum) are unambiguously shown already at 50 days. The spectra 
remain surprisingly blue and the blackbody temperature fits to the photometric flux shows a lack of cooling over 300 days.
The density (n$_{\rm e}\sim10^6$ cm$^{-3}$) and line velocity (v$\sim3000$ \kms) of these oxygen features are different than the bulk of ejecta (n$_{\rm e}\sim10^9$ cm$^{-3}$, v$\sim$8000 \kms), suggesting that multiple emitting regions are responsible for the features observed in the spectra.  This could be caused by an axis-symmetric or fragmented ejecta or by an additional component arising in a circumstellar material.
Furthermore, an in depth analysis of the light curve shows hints of undulations as observed in other similar SLSNe~I (e.g. SN~2015bn), as well a decline faster than \co\/ after 150 days and even faster after 300d (times measured from estimated maximum light). We investigated X-ray upper limits, presenting new limits for \an\/ and SN~2015bn, and spectrophotometric behaviour with those of nearby slow-evolving SLSNe~I showing similar coverage. The deepest limits (not stacked) disfavour the circumstellar interaction with a massive CSM ($\sim$13 \M) that is usually needed to reproduce the slow-evolving SLSNe light curves.

We noticed that some of these observational features are not unique to \an\/ but are common in, or even exclusive to, all slow-evolving SLSNe~I observed to date 
\begin{itemize}
\item~they show semi-forbidden and forbidden lines from 50-70 days to peak with little  to no variation up to 400 days.
The strongest evolution is seen in the strengths of the [O~{\sc i}] $\lambda\lambda$6300, 6363 lines. 
The centroid of oxygen and calcium lines appear to move in time. In addition, ionised elements show line profiles different from the neutral ones and likely from a region interior to that where the neutral lines are formed, and where occultation effects can be easily produced. This suggests that multiple emitting regions are responsible of the overall spectra. 
\item~[Ca~{\sc ii}] appears earlier than in broad-line type Ic SNe and fast-evolving SLSNe. In two cases (SN~2015bn and \an\/) it exhibits a blue-shifted  peak.  In 
 \an\/ the line is likely dominated by a [O~{\sc ii}] component. The Ca~{\sc ii} NIR triplet is observed increasing with time, from 100 days onward, in contrast to what shown in normal stripped envelope SNe. 
\item the overall light curve evolution of these four objects are similar. Initially, all of them decay at a rate consistent with 
  \co\/ powering. But after 150 days their bolometric light curves decline faster than the fully trapped \co\/ to \fe\/ decay.
Moreover, after $\sim$300 days from maximum light the bolometric light curves show a further steepening in the decline consistent with a power law of $t^{-5}$.  
The light curves overall are not consistent with 
massive ejecta from PISNe that fully trap the gamma rays from \co\/ decay up to 500 days.
\item light curve fluctuations seem to be common in the slow-evolving. They can be generally explained with changes in the density structure of the expanding material. Since they appear to 
occur more than once and at different phases in each object, they are likely due to interaction with a clumpy medium. The medium should have M$<$0.04 \M\/ to explain the oscillations observed in four out of five inspected slow-evolving SLSNe~I and to account for the X-ray limits here reported. 

\end{itemize} 

There remains a challenge to explain all these  observables, together with those previously reported to date. 
For example the observed  pre peak bumps, axis-symmetric ejecta, late spectra similar to broad line type Ic and reproducible by $\sim10$\M\/ of oxygen \citep{ni15a,sm16,ni16a1,in16c,je16,ni16b}, with a common scenario. The two most appealing mechanisms 
for these slowly evolving SLSNe are still central engine (likely a magnetar) and the interaction, even though for the latter some observables are not in agreement with the classical observed behaviour in interacting transients. However, some further considerations can be made. At least a small degree of interaction seems to happen in slow-evolving SLSNe~I. If the slow and fast evolving SLSNe~I share the same nature (i. e. mechanism and maybe progenitor system), the difference between them cannot be due exclusively to a difference in mass as previously suggested \citep{ni15} since that would not reflect the spectral evolution.

To understand if fast and slow-evolving SLSNe~I are a similar or different kind of transients we need further, and more statistical, investigations. Further studies should also focus on a more detailed early (when we have available both the types) spectra analysis and modelling in order to increase the available information. 

\section*{Acknowledgments}
CI thanks Ragnhild Lunnan for sharing the data of PS1-14bj and Stuart Sim for stimulating discussions.
This work is based on observations collected at the European Organisation for Astronomical Research in the Southern Hemisphere, Chile as part of PESSTO, (the Public ESO Spectroscopic Survey for Transient Objects Survey) ESO program 188.D-3003, 191.D-0935, 197.D-1075, and the X-Shooter programmes 093.D-0229, 092.D-0555. 
The research leading to these results has received funding from the European Research Council under the European Union's Seventh Framework Programme (FP7/2007-2013)/ERC Grant agreement no [291222]. SJS acknowledges funding from STFC grants ST/I001123/1 and ST/L000709/1. C.I. thanks the organisers and participants of the Munich Institute for Astro- and Particle Physics (MIAPP) workshop "Supernovae: The Outliers" for stimulating discussions. This research was supported by the Munich Institute for Astro- and Particle Physics (MIAPP) of the DFG cluster of excellence "Origin and Structure of the Universe". MF acknowledges the support of a Royal Society - Science Foundation Ireland University Research Fellowship. KM acknowledges support from the STFC through an Ernest Rutherford Fellowship.  TWC and TK acknowledge the support through the Sofia Kovalevskaja Award to P. Schady from the Alexander von Humboldt Foundation of Germany. The Liverpool Telescope is operated on the island of La Palma by Liverpool John Moores University in the Spanish Observatorio del Roque de los Muchachos of the Instituto de Astrofisica de Canarias with financial support from the UK Science and Technology Facilities Council. The Pan-STARRS1 Surveys (PS1) have been made possible through contributions of the Institute for Astronomy, the University of Hawaii, the Pan-STARRS Project Office, the Max-Planck Society and its participating institutes, the Max Planck Institute for Astronomy, Heidelberg and the Max Planck Institute for Extraterrestrial Physics, Garching, The Johns Hopkins University, Durham University, the University of Edinburgh, Queen's University Belfast, the Harvard-Smithsonian Center for Astrophysics, the Las Cumbres Observatory Global Telescope Network Incorporated, the National Central University of Taiwan, the Space Telescope Science Institute, the National Aeronautics and Space Administration under Grant No. NNX08AR22G issued through the Planetary Science Division of the NASA Science Mission Directorate, the National Science Foundation under Grant No. AST-1238877, the University of Maryland, and Eotvos Lorand University (ELTE) and the Los Alamos National Laboratory.

\appendix

\section{Tables}

\begin{table*}
\caption{Journal of spectroscopic observations of \an.}
\begin{center}
\begin{tabular}{ccccccl}
\hline
Date & MJD & Phase* &Range (observer-frame) & Range (rest-frame) &Resolution & Instrumental\\
yy/mm/dd &  & (days)  &(\AA) &(\AA)  & (\AA) &Configuration\\
\hline
14/01/02 &       56660.31 & 55.0 & 3665--9285 &3150--7980 & 18 & NTT+EFOSC2+gm13\\                  
14/01/08&        56665.81        & 59.7  & 3360--10010 & 2890--8600  & 14/13 & NTT+EFOSC2+gm11/16         \\  
14/01/29&       56687.28        & 78.2  & 3095--24785 &2660--21300   & 1.0/1.1/3.3 & VLT+XSHOOTER+UV/OPT/NIR \\                                                                                  
14/01/29&       56687.31         &  78.2  &  3360--10010 & 2890--8600  &  14/13   & NTT+EFOSC2+gm11/16      \\                                                                                                                                                          
14/02/21&        56710.70 & 98.3 &  3360--10010 &2890--8600 &  14/13  & NTT+EFOSC2+gm11/16  \\                       
14/03/08  &      56725.16 & 110.7& 3665--9285 &3150--7980&   18 & NTT+EFOSC2+gm13\\                                                       
14/04/22&         56770.02  &149.3& 3665--9285 &3150--7980&18 & NTT+EFOSC2+gm13\\                                                                                        
14/06/21 &       56829.09&  200.0 & 3095--24785 &2660--21300  & 1.0/1.1/3.3  & VLT+XSHOOTER+UV/OPT/NIR\\                                      
14/08/21 &       56890.49 & 252.8 & 3665--9285 &3150--7980   & 18 & NTT+EFOSC2+gm13\\                                                                                                                                       
\hline
\end{tabular}
\end{center}
* Phase with respect to the assumed maximum.
\label{table:sp}
\end{table*}%

\begin{table*}
\caption{$gri$ light curves of \an.}
\begin{center}
\begin{tabular}{ccccccc}
\hline
Date & MJD & Phase$^*$ & {\it g} & {\it r} & {\it i} & Instrument$^{\dagger}$\\
yy/mm/dd &  & (days)   & & & &\\
\hline
14/01/02	&	56659.28	&	54.1	&		&	18.60 (0.08)	&		&	LSQ \\
14/01/04	&	56661.28	&	55.8	&		&	18.62 (0.08)	&		&	LSQ \\
14/01/06	&	56663.30	&	57.6	&		&	18.64 (0.08)	&		&	LSQ \\
14/01/14	&	56672.19	&	65.2	&	18.93 (0.13)	&	18.92 (0.19)	&	18.89 (0.17)	&	LT \\
14/01/28	&	56686.23	&	77.3	&	19.15 (0.11)	&	19.12 (0.11)	&	18.99 (0.12)	&	LT \\
14/02/01	&	56689.24	&	79.9	&	19.12 (0.13)	&	19.13 (0.14)	&	18.99 (0.12)	&	NTT \\
14/02/07	&	56696.24	&	85.9	&	19.11 (0.17)	&	19.14 (0.11)	&	19.09 (0.12)	&	LT \\
14/02/13	&	56702.11	&	90.9	&	19.19 (0.16)	&	19.13 (0.14)	&	19.24 (0.15)	&	LT \\
14/03/12	&	56729.03	&	114.1	&	19.35 (0.18)	&	19.36 (0.14)	&	19.40 (0.16)	&	LT \\
14/04/21	&	56769.17	&	148.6	&	19.68 (0.12)	&	20.02 (0.14)	&	19.83 (0.12)	&	NTT \\
14/05/16	&	56793.93	&	169.8	&	20.06 (0.14)	&	20.13 (0.14)	&	20.13 (0.14)	&	LT \\
14/05/19	&	56796.89	&	172.4	&	20.03 (0.12)	&	20.19 (0.12)	&	20.18 (0.13)	&	LT \\
14/05/23	&	56800.88	&	175.8	&	20.11 (0.12)	&	20.32 (0.12)	&	20.45 (0.14)	&	LT \\
14/05/27	&	56804.91	&	179.3	&	20.07 (0.11)	&	20.41 (0.12)	&	20.49 (0.14)	&	LT \\
14/06/03	&	56811.96	&	185.3	&	20.07 (0.12)	&	20.46 (0.13)	&	20.55 (0.15)	&	LT \\
14/08/25	&	56894.99	&	256.7	&	21.55 (0.11)	&	21.46 (0.14)	&	21.60 (0.23)	&	NTT \\
14/12/29	&	57021.30	&	365.2	&	22.50 (0.13)	&	22.49 (0.15)	&	22.82 (0.19)	&	NTT \\
15/01/26	&	57049.30	&	389.3	&	22.98 (0.15)	&	23.03 (0.19)	&	23.18 (0.25)	&	NTT \\
15/02/17	&	57071.20	&	408.1	&	23.64 (0.19)	&	23.83 (0.22)	&	24.06 (0.32)	&	NTT \\
\hline
\end{tabular}
\end{center}
* Phase with respect to the assumed maximum

$\dagger$ LSQ = La Silla Quest; LT = Liverpool Telescope + IO:O; NTT = ESO NTT + EFOSC2
\label{table:snm}
\end{table*}%

\begin{table*}
\caption{$BVRJHK$ light curve of LSQ14an measured after template subtraction (from 2015-04-17) using HOTPANTS}
\begin{center}
\begin{tabular}{ccccccc}
\hline
Date & MJD & Phase$^*$ & $B$& $V$ & $R$ & Instrument$^{\dagger}$\\
yy/mm/dd &  & (days)  &   & & &\\
\hline
14/01/02 &  56660.31 & 55.0 &&   19.00 (0.10) &  &     NTT\\
14/01/08 & 56665.81 &  59.7  &&  19.10 (0.13) &  &     NTT\\
14/01/29 & 56687.28 &  78.2  &&  19.49 (0.09) &  &    NTT\\
14/02/21 & 56710.70 &  98.3  &&  19.63 (0.12) &  &     NTT\\
14/03/08 & 56725.16 &  110.7&  &  19.68 (0.18) &  &     NTT\\
14/04/22 & 56770.02 &  149.3 & &  20.13 (0.14) &  &     NTT\\
14/08/25	&	56894.99	&	256.7	&	21.85 (0.10)	&	21.50 (0.03)	&	21.31 (0.09)		&	NTT \\
14/12/29	&	57021.30	&	365.2	&	22.72 (0.15)	&	22.47 (0.06)	&	22.39 (0.10)		&	NTT \\
15/01/26	&	57049.30	&	389.3	&	23.21 (0.13)	&	22.99 (0.10)	&	22.93 (0.15)		&	NTT \\
15/02/17	&	57071.20	&	408.1	&	23.95 (0.20)	&	23.46 (0.16)	&	23.84 (0.19)		&	NTT \\
\hline
Date & MJD & Phase$^*$ & $J$ & $H$ & $K$ & Instrument$^{\dagger}$\\
yy/mm/dd &  & (days)  &   & & &\\
\hline
14/04/23 &  56771.04 & 150.16 &    19.93  (0.12) &    20.04 (0.12)  &   20.08  (0.20) &  NTT\\
\hline
\end{tabular}
\end{center}
* Phase with respect to the assumed maximum
$\dagger$  NTT = ESO NTT + EFOSC2
\label{table:snmv}
\end{table*}%

\begin{table*}
\caption{Magnitudes in $griBVR$ of the local sequence stars in the field of \an.}
\begin{center}
\begin{tabular}{ccccc}
\hline
RA& Dec & g	&	r	&	i\\
\hline
12:53:52.095  &	-29:32:20.47	& 16.36	(0.05) & 	15.92	(0.05) &	15.74	(0.05)\\
12:53:49.806  &	-29:31:58.25	& 17.61	(0.07) & 	16.37	(0.05) &	15.56	(0.05)\\
12:53:49.136  &	-29:32:15.38	& 17.11	(0.05) & 	16.61	(0.04) &	16.38	(0.04)\\
12:53:48.019  &	-29:31:37.48	& 17.78	(0.07) & 	16.51	(0.04) &	15.67	(0.05)\\	
12:53:44.110  &		-29:32:04.44	& 16.76	(0.06) & 	16.32	(0.05) &	16.13	(0.05)\\
12:53:52.622  &	-29:29:23.17	& 16.38	(0.05) & 	15.93	(0.04) &	15.75	(0.05)\\
12:53:52.300  &	-29:29:36.81	& 16.22	(0.05) & 	15.59	(0.04) &	15.27	(0.05)\\
12:53:44.595  &	-29:29:30.65	& 17.76	(0.08) & 	17.22	(0.04) &	16.98	(0.03)\\
\hline
RA& Dec & B	&	V	&	R\\
\hline
12:53:52.672 &	-29:30:12.65 &	20.56	(0.13) &	19.60	(0.08) &	18.95	(0.07)	\\
12:53:40.273 &	-29:30:15.14 &	19.35	(0.12) &	18.52	(0.10) &	18.02	(0.10)	\\
12:53:40.049 &	-29:30:54.67 &	18.93	(0.11) &	18.22	(0.09) &	17.76	(0.09)	\\
12:53:44.735 &	-29:31:10.91 &	20.88	(0.12) &	20.41	(0.10) &	20.04	(0.08)	\\
12:53:47.986 &	-29:31:37.54 &	18.62	(0.14) &	17.16	(0.10) &	16.19	(0.08)	\\
12:53:49.102 &	-29:32:15.83 &	17.67	(0.13) &	16.95	(0.10) &	16.48	(0.08)	\\
12:53:55.383 &	-29:32:04.58 &	20.59	(0.14) &	20.00	(0.10) &	19.62	(0.09)	\\
\hline
\end{tabular}
\end{center}
\label{table:sref}
\end{table*}%

\begin{table*}
\caption{Log of X-ray observations of LSQ14an and SN~2015bn}
\begin{center}
\begin{tabular}{cccccc}
\hline
\hline
\multicolumn{6}{c}{LSQ14an}\\
\hline
Date & MJD & Phase$^*$ & Exposure& F$_{\rm X}$& L$_{\rm X}$ \\
yy/mm/dd &  & (days)  & (s)  & (erg cm$^{-2}$ s$^{-1}$)& (erg s$^{-1}$)\\
\hline
14/03/24 & 56740.52  & 123.9 &9000&  $<$1.0e-13   &  $<$7.0e42   \\
14/07/03 & 56841.04  &  210.3  &5400&  $<$2.6e-13 &   $<$1.8e43  \\
14/12/07 & 56998.41  &  345.5  &7600&    $<$1.8e-13 & $<$1.3e43 \\
16/08/16 & 57616.01  &  876.3  &9500&    $<$5.6e-14 & $<$3.9e42 \\
\hline
Stacked &  &  & & & \\
\hline
 & 56859.70$^{\dagger}$ & 226.3$^{\dagger}$ & 22000 & $<$4.1e-14 & $<$2.9e42 \\
\hline
\hline
\multicolumn{6}{c}{SN~2015bn}\\
\hline
Date & MJD & Phase$^*$ & Exposure& F$_{\rm X}$& L$_{\rm X}$ \\
yy/mm/dd &  & (days)  & (s)  & (erg cm$^{-2}$ s$^{-1}$)& (erg s$^{-1}$)\\
\hline
15/02/19 & 57072.29 & -26.7 &6000&   $<$2.7e-13   &  $<$8.6e42  \\
15/04/19 & 57131.26  & 26.3 &6700&   $<$8.7e-14 &  $<$2.8e42 \\
16/01/03 & 57390.02  & 258.6 & 8700&  $<$3.6e-14 &  $<$1.1e42 \\
16/03/02 & 57449.07 & 311.6  & 2800&   $<$3.7e-13 & $<$1.2e43 \\
16/07/04 & 57573.21 & 423.1 & 3900& $<$3.2e-13  &  $<$1.0e43\\
16/12/16 & 57738.87& 571.8 & 8900 & $<$3.6e-14 & $<$1.1e42 \\
16/12/31 & 57753.58 & 585.0  & 9940 & $<$3.0e-14 & $<$9.5e41 \\
17/01/19 & 57772.01& 601.5 & 9920 &  $<$3.2e-14 & $<$1.0e42\\
17/02/06 & 57788.56 & 616.5 & 8950 & $<$3.5e-14 & $<$1.1e42 \\
17/02/17 & 57801.71 & 628.3 & 8900 & $<$5.0e-14 & $<$1.6e42 \\
\hline
Stacked &  &  & & & \\
\hline
  & 57323.17$^{\ddagger}$ &198.6$^{\ddagger}$ & 28000 & $<$2.4e-14 & $<$7.6e41  \\
 \hline
\end{tabular}
\label{table:xr}
\end{center}
* Phase with respect to the assumed maximum

$\dagger$ Average date/MJD between the first three exposures
$\dagger\dagger$ Average date/MJD between the first four exposures
\label{table:snx}
\end{table*}

\begin{table*}
\begin{center}
\begin{tabular}[t]{lllllll}
\hline 
\hline
Line & $\lambda$ & Observed flux & Error & RMS & EW & FWHM \\
 & (\AA) & ($10^{-17}$ erg s$^{-1}$ cm$^{-2}$) & ($10^{-17}$ erg s$^{-1}$ cm$^{-2}$) & ($10^{-17}$ erg s$^{-1}$ cm$^{-2}$) & (\AA)  & (\AA) \\
\hline
He~{\sc i}&3188&3.92&1.98&0.52&0.55&1.88\\
He~{\sc i}&3704&1.43&0.91&0.29&0.20&1.29\\
\oii&3726&65.65&2.39&0.42&8.94&1.64\\
\oii&3729&92.03&2.69&0.42&12.46&1.63\\
H12&3750&1.22&1.03&0.34&0.16&1.24\\
H11&3770&1.70&0.98&0.32&0.23&1.24\\
H10&3798&2.89&0.73&0.23&0.43&1.28\\
H9&3835&6.79&1.44&0.37&0.91&1.76\\
\neiii&3868&34.63&1.76&0.37&4.51&1.54\\
He~{\sc i}+H8&3889&12.27&0.96&0.25&1.65&1.50\\
\neiii&3967&7.40&0.93&0.27&0.94&1.36\\
H7&3969&9.65&0.95&0.27&1.23&1.36\\
\nii+He~{\sc i}&4026&0.82&0.44&0.16&0.12&1.00\\
H$\delta$&4102&17.14&0.90&0.22&2.88&1.54\\
H$\gamma$&4340&35.26&1.29&0.24&5.16&1.86\\
\oiii&4363&5.43&1.03&0.30&0.82&1.86\\
He~{\sc i}&4471&4.35&0.87&0.20&0.86&2.44\\
\Hb&4861&68.47&2.06&0.24&22.68&2.03\\
\oiii&4959&129.20&3.10&0.37&23.43&1.96\\
\oiii&5007&345.70&4.65&0.34&68.15&2.03\\
He~{\sc i}&5876&7.75&0.57&0.11&4.90&2.25\\
\oi&6300&4.88&0.96&0.19&2.12&2.90\\
\siii&6312&1.56&0.85&0.18&0.65&2.90\\
\oi&6364&1.36&0.43&0.11&0.51&2.15\\
\nii&6548&2.11&1.61&0.35&1.01&2.73\\
\Ha&6563&234.30&3.37&0.20&105.19&2.73\\
\nii&6584&5.25&0.54&0.11&2.67&2.73\\
He~{\sc i}&6678&1.96&0.33&0.07&1.43&2.72\\
\sii&6717&13.83&0.78&0.11&10.87&2.74\\
\sii&6731&10.21&0.68&0.11&8.19&2.74\\
He~{\sc i}&7065&1.87&0.41&0.09&1.36&2.75\\
\ariii&7136&4.89&1.16&0.22&3.10&2.82\\
\oii&7320&0.95&0.51&0.12&0.19&2.47\\
\oii&7330&1.05&0.52&0.12&0.21&2.47\\
\siii&9069&15.74&1.05&0.15&10.96&3.28\\
\siii&9532&36.59&2.86&0.31&24.45&3.42\\
\hline
\end{tabular}
\caption{Observed emission line measurements of the host galaxy of LSQ14an from the spectrum taken by VLT + Xshooter at +200.0d (no applying any reddening corrections). The EW of each line is a minimum, not a real line strength due to the SN contamination of the continuum.} 
\label{tab:other_14an_flux_vlt}
\end{center}
\end{table*}

\end{document}